\documentstyle[preprint,revtex]{aps}	
\begin{document}
\topmargin 0.7in
\oddsidemargin 0.8in
\evensidemargin 0.8in
\draft
%
%
\preprint{FSU-HEP-930519}
\preprint{FERMILAB-PUB-93/096-T}
\preprint{DTP/93/26}
\preprint{May 1993}
\begin{title}
QCD Corrections to Hadronic $W\gamma$ Production\\
with Non-standard $WW\gamma$ Couplings
\end{title}
\author{U.~Baur}
\begin{instit}
Department of Physics, Florida State University, Tallahassee, FL 32306, USA
\end{instit}
\vskip -2.mm
\author{T.~Han}
\begin{instit}
Fermi National Accelerator Laboratory, P.O. Box 500, Batavia, IL 60510, USA
\end{instit}
\vskip -2.mm
\author{J.~Ohnemus}
\begin{instit}
Department of Physics, University of Durham, Durham, DH1 3LE, England
\end{instit}
\thispagestyle{empty}
\vskip -8.mm
\begin{abstract}
\nonum\section{abstract}
\baselineskip16.pt  
The process $p\,p\hskip-7pt\hbox{$^{^{(\!-\!)}}$} \rightarrow
W^{\pm}\gamma + X \rightarrow \ell^\pm \nu \gamma + X$ is calculated to
${\cal O}(\alpha_s)$ for general $CP$ conserving $WW\gamma$ couplings.
At the Tevatron center of mass energy, the QCD corrections to $W\gamma$
production are modest, and the Born and inclusive ${\cal O}(\alpha_s)$
cross sections have similar sensitivities to the effects of
anomalous couplings. At supercollider energies, the inclusive QCD
corrections are large at high photon transverse momenta, reducing the
sensitivity to non-standard $WW\gamma$ couplings by up to a factor~2.
The size of the QCD corrections can be reduced significantly, and a large
fraction of the sensitivity lost can be regained, if a jet veto is imposed.
\end{abstract}
\newpage
%
%
\begin{narrowtext}
\section{INTRODUCTION}

The electroweak standard model (SM) based on an
$\hbox{\rm SU(2)} \bigotimes \hbox{\rm U(1)}$ gauge theory
has been remarkably successful in describing contemporary high
energy physics experiments. The three vector boson couplings
predicted by this non-abelian gauge theory, however, remain largely untested.
The production of $W\gamma$ pairs at hadron colliders provides an
excellent opportunity to study the $WW\gamma$ vertex~\cite{BROWN,EHLQ}.
In addition, the reaction $p\,p\hskip-7pt\hbox{$^{^{(\!-\!)}}$}
\rightarrow W^{\pm}\gamma$ is of special interest due to the presence
of a zero in the amplitude of the
parton level subprocess $q_1 \bar q_2 \rightarrow W \gamma$~\cite{RAZ}.
This phenomenon may make it possible to measure the magnetic dipole
moment and electric quadrupole moment of the $W$-boson~\cite{VALENZUELA,CHH}.
In the SM, the $WW\gamma$ vertex is completely fixed by the
$\hbox{\rm SU(2)} \bigotimes \hbox{\rm U(1)}$ gauge structure of the
electroweak sector. A measurement of the $WW\gamma$ vertex thus provides
a stringent test of the SM.

In contrast to low energy data and high precision measurements at the $Z$
peak, collider experiments offer the possibility of a direct, and
essentially model independent, determination of the three vector boson
vertices. Hadronic production of $W\gamma$ pairs was first calculated in
Ref.~\cite{BROWN}. The ${\cal O}(\alpha_s)$ QCD corrections to the reaction
$p\,p\hskip-7pt\hbox{$^{^{(\!-\!)}}$} \rightarrow W^{\pm}\gamma$
were first evaluated in Ref.~\cite{SMITH}.
${\cal O}(\alpha_s^2)$ QCD corrections in the soft-plus-virtual gluon
approximation were recently estimated in Ref.~\cite{MENDOZA}.
Studies on the potential for probing the $WW\gamma$
vertex have been performed for $e^+ e^-$~\cite{WGAMMAEE},
$ep$~\cite{WGAMMAEP}, and
$p\,p\hskip-7pt\hbox{$^{^{(\!-\!)}}$}$ \cite{CHH,BAURZEP,BAURBERGER,LHCWORK}
collisions.  A general discussion of non-standard model couplings of the
$W$-boson has been given in Ref.~\cite{BAURZEP}.
The first experimental observation of $W\gamma$ production in hadronic
collisions has recently been reported by the UA2 Collaboration~\cite{UATWO}.

Previous studies on probing the $WW\gamma$ vertex via hadronic $W\gamma$
production have been based on leading-order (LO)
calculations \cite{CHH,BAURZEP,BAURBERGER,LHCWORK}.  In general, the
inclusion of anomalous couplings at the $WW\gamma$ vertex yields
enhancements in the $W\gamma$ cross section, especially at large values of
the photon transverse momentum, $p_T^{}(\gamma)$,
and at large values of the $W\gamma$
invariant mass, $M_{W\gamma}$ \cite{CHH,BAURZEP,BAURBERGER,LHCWORK}.
A recent next-to-leading-order (NLO) calculation of hadronic $W\gamma$
production~\cite{WGAMMA} has shown that the ${\cal O}(\alpha_s)$ corrections
are large in precisely these same regions.
Furthermore, higher order corrections and anomalous couplings
both destroy the amplitude zero of the lowest order process.
It is thus vital to include the NLO corrections when using
hadronic $W\gamma$ production to test the $WW\gamma$ vertex for
anomalous couplings.

In this paper, we calculate hadronic $W\gamma$ production to ${\cal
O}(\alpha_s)$, including the most general, $CP$ conserving, anomalous
$WW\gamma$ couplings. We also include the leptonic decay
of the $W$-boson in the narrow width approximation in our
calculation.  In this approximation, diagrams where the photon is
radiated off the final state lepton line are not necessary to maintain
electromagnetic gauge invariance. For suitable cuts these diagrams can
thus be ignored, which considerably simplifies the calculation.
Our calculation, which has been
performed using the Monte Carlo method for NLO calculations~\cite{NLOMC}, is
described in Section~II. With this method, it
is easy to calculate a variety of observables simultaneously and to
implement experimental acceptance cuts in the calculation. It is also
possible to compute the NLO QCD corrections for exclusive channels,
{\it e.g.}, $p\,p\hskip-7pt\hbox{$^{^{(\!-\!)}}$} \rightarrow W\gamma
+0$~jet. Apart from anomalous contributions to the $WW\gamma$ vertex we
assume the SM to be valid in our calculation. In particular, we assume
the coupling of the $W$ bosons to quarks and leptons to be given by the SM.

The results of our numerical simulations are given in Section~III. At
supercollider energies, the inclusive NLO QCD corrections are very large
at high photon transverse momenta in the SM. They have a
severe negative impact on the sensitivity bounds for anomalous
$WW\gamma$ couplings which can be achieved at the SSC or LHC.
The large QCD corrections are caused by
the combined effects of destructive interference in the Born subprocess,
a log squared enhancement factor in the $q_1 g \rightarrow W \gamma q_2$
partonic cross section at high photon
transverse momentum \cite{FRIX}, and the large quark-gluon luminosity
at supercollider energies. At the Tevatron, on the other hand, the
${\cal O}(\alpha_s)$ QCD corrections are found to be modest and
sensitivities are only slightly affected by the QCD corrections.
In Section~III, we also show that the QCD corrections at high
$p_T^{}(\gamma)$ can be significantly reduced and a large fraction of the
sensitivity to anomalous couplings lost at supercollider energies can
be regained, if a jet veto is imposed, {\it i.e.}, if the $W\gamma+0$~jet
exclusive channel is used to extract information on the $WW\gamma$
vertex. We also find that the residual dependence of the NLO
$W\gamma+0$~jet cross section on the factorization scale
$Q^2$ is significantly smaller than that of the ${\cal O}(\alpha_s)$
cross section for the inclusive reaction $p\,p\hskip-7pt
\hbox{$^{^{(\!-\!)}}$} \rightarrow W\gamma+X$. Our conclusions are given
in Section~IV. Finally, there are two appendices containing
technical details of the calculation.

\section{FORMALISM}

An ${\cal O}(\alpha_s)$ calculation of hadronic $W\gamma$ production
was recently presented in Ref.~\cite{WGAMMA}.  The calculation was performed
for a real
$W$-boson in the final state and assumed all couplings had their standard
model values.  The results of Ref.~\cite{WGAMMA} are extended in this section
to include the leptonic decay $W \to \ell \nu$ $(\ell = e, \mu)$
and anomalous (non-standard model) couplings at the $WW\gamma$ vertex.
First, the NLO Monte Carlo formalism used in this calculation is
summarized and the results of Ref.~\cite{WGAMMA} are outlined.
These results are then generalized to include the decay $W \to \ell
\nu$ and the most general $CP$ conserving $WW\gamma$ couplings.

The calculation is done using the narrow width approximation for
the $W$ decay. This simplifies the calculation greatly for two
reasons.  First of all, it is possible to ignore Feynman diagrams in which
the photon is radiated off the final state lepton line without
violating electromagnetic gauge invariance.  (Radiative $W$ decay
events can be suppressed by a suitable choice of cuts~\cite{BAURBERGER}
which we will impose in our numerical simulations; see Section~IIIB.)
Secondly, in the narrow width approximation it is particularly easy to extend
the NLO calculation of Ref.~\cite{WGAMMA} to include the leptonic decay of the
$W$-boson.

\subsection{Monte Carlo Formalism}

The NLO calculation of $W\gamma$ production includes contributions from the
square of the Born graphs shown in Fig.~\ref{FIG:BORNGRAPHS}, the interference
between the Born graphs and the virtual one-loop graphs shown in
Fig.~\ref{FIG:VIRTUALGRAPHS}, and the square of the real emission graphs shown
in Fig.~\ref{FIG:REALGRAPHS}. Our calculation has been carried out
using a combination of analytic and Monte Carlo integration
methods~\cite{NLOMC}. The basic idea is to isolate the
soft and collinear singularities associated with the real emission
subprocesses by partitioning phase space into soft, collinear, and
finite regions.  This is done by introducing theoretical soft and
collinear cutoff parameters, $\delta_s$ and $\delta_c$.  Using
dimensional regularization~\cite{TV},  the soft and collinear
singularities are exposed as poles in  $\epsilon$ (the number of
space-time dimensions is $N = 4 - 2\epsilon$ with $\epsilon$ a small
number).  The infrared singularities from the soft and virtual
contributions are then explicitly canceled while the collinear
singularities are factorized and absorbed into the definition of the
parton distribution functions or the photon fragmentation functions.
The remaining contributions are finite
and can be  evaluated in four dimensions.  The Monte Carlo program thus
generates $n$-body (for the Born and virtual contributions) and
$(n+1)$-body (for the real emission contributions) final state events.
The $n$- and $(n+1)$-body contributions both depend on the cutoff
parameters $\delta_s$ and $\delta_c$, however, when these contributions
are added together to form a suitably inclusive observable, all
dependence on the cutoff parameters cancels.
The numerical results presented in this paper are insensitive to
variations of the cutoff parameters; this will be demonstrated later.

\subsection{Summary of ${\cal O}(\alpha_s)$ $W\gamma$ Production }

The NLO cross section for hadronic $W\gamma$ production~\cite{WGAMMA}
consists of two- and three-body final state contributions:
\widetext
\FL
\begin{eqnarray}
\noalign{\vskip 5pt}
\sigma^{\hbox{\scriptsize NLO}} ( p\,p\hskip-7pt\hbox{$^{^{(\!-\!)}}$}
\rightarrow W\gamma + X) =
\sigma^{\hbox{\scriptsize NLO}}_{\hbox{\scriptsize 2 body}}
(p\,p\hskip-7pt\hbox{$^{^{(\!-\!)}}$} \rightarrow W\gamma)
+ \sigma_{\hbox{\scriptsize 3 body}} (p\,p\hskip-7pt\hbox{$^{^{(\!-\!)
}}$} \rightarrow W\gamma + X) \>.
\label{EQ:NLOCROSSW}
\end{eqnarray}
\narrowtext
The two-body contribution is
\widetext
\FL
\begin{eqnarray}
\noalign{\vskip 5pt}
& &\sigma^{\hbox {\scriptsize NLO}}_{\hbox{\scriptsize 2 body}}
 (p\,p\hskip-7pt\hbox{$^{^{(\!-\!)}}$} \rightarrow W\gamma) =
\sigma^{\hbox{\scriptsize NLO}}_{\hbox{\scriptsize brem}}
 + \sigma^{\hbox{\scriptsize hc}}
 + \sum_{q_1, \bar q_2} \int dv\,dx_1\,dx_2  \label{EQ:NLOTWO} \\
\noalign{\vskip 5pt}
 & &\quad \times \biggl[ G_{q_1/p}(x_1,M^2)\,G_{\bar q_2/p\hskip-7pt
\hbox{$^{^{(\!-\!)}}$}}(x_2,M^2) \,
 {d\hat\sigma^{{\hbox {\scriptsize NLO}}}\over dv}
 (q_1\bar q_2\rightarrow W \gamma)
 + (x_1 \leftrightarrow x_2) \biggr] \>,
\nonumber
\end{eqnarray}
\narrowtext
where the quantities $\sigma^{\hbox{\scriptsize NLO}}_{\hbox{\scriptsize
brem}}$ and $\sigma^{\hbox{\scriptsize hc}}$
are the contributions from the NLO
bremsstrahlung cross section and the hard collinear remnants, respectively.
These contributions are defined in Appendices~A and B, respectively, for the
case of $W\gamma$ production with leptonic decay of the $W$-boson.
In Eq.~(\ref{EQ:NLOTWO}), the sum is over all contributing quark flavors,
$v$ is related to the center of
mass scattering angle $\theta^*$ by $v = {1\over 2}(1 + \cos\theta^*)$,
$x_1$ and $x_2$ are the parton momentum fractions, $G_{q/p}(x,M^2)$ is a parton
distribution function, $M^2$ is the factorization scale, and
\FL
\begin{eqnarray}
\noalign{\vskip 5pt}
{d\hat\sigma^{{\hbox {\scriptsize NLO}}}\over dv}
 (q_1\bar q_2 \rightarrow W\gamma) = &\phantom{+}&
 {d\hat\sigma^{\hbox{\scriptsize Born}}\over dv} (q_1 \bar q_2 \to W\gamma) \,
 \Biggl[ 1 + C_F {\alpha_s(\mu^2) \over 2 \pi} \biggl\{ 4 \ln(\delta_s)^2
+3\ln\Bigl({s\over M^2}\Bigr) \nonumber \\
\noalign{\vskip 5pt}
&+& 4 \ln(\delta_s) \ln\Bigl({s\over M^2}\Bigr)
+ \lambda_{FC} \Bigl( 9 + {2\over 3} \pi^2 + 3 \ln(\delta_s)
- 2 \ln(\delta_s)^2 \Bigr) \biggr\} \Biggl] \nonumber \\
\noalign{\vskip 5pt}
&+& {d\hat\sigma^{\hbox{\scriptsize virt}}\over dv}
(q_1 \bar q_2 \to W\gamma) \>.
\label{EQ:NLO}
\end{eqnarray}
Here $C_F = {4\over 3}$ is the quark-gluon vertex color factor,
$\alpha_s(\mu^2)$ is the strong running coupling evaluated at
the renormalization scale $\mu^2$, $\delta_s$ is the soft cutoff
parameter, and $\lambda_{FC}$
specifies the factorization convention: $\lambda_{FC} = 0$ for the universal
(Modified Minimal Subtraction ${\rm \overline{MS}}$~\cite{MSBAR}) convention
and $\lambda_{FC} = 1$ for the physical (Deep Inelastic Scattering DIS)
convention.

The ${\cal O}(\alpha_s)$ virtual contribution to the $q_1 \bar q_2
\rightarrow W \gamma$ cross section is
\FL
\begin{eqnarray}
\noalign{\vskip 5pt}
{d\hat\sigma^{\hbox{\scriptsize virt}} \over dv} (q_1 \bar q_2 \to W\gamma) =
&\phantom{+}& C_F \, {\alpha_s(\mu^2) \over 2 \pi} \,
{1\over 4} {1\over 9}\, {(s-M_W^2) \over 16 \pi s^2}\,
N_C \, e^4 \, { |U_{q_1 q_2}|^2 \over 2 x_{\hbox{\scriptsize w}} }
\label{EQ:VIRT} \\
\noalign{\vskip 5pt}
&\times& {(Q_1 t + Q_2 u) \over t+u} \,
\biggl[ Q_1 F^{W}(t,u) +  Q_2 F^{W}(u,t) \biggl] \>,
\nonumber
\end{eqnarray}
where
\FL
\begin{eqnarray}
F^{W}(t,u) = &\phantom{+}& 4 \Bigl[ 2{s^2\over tu} + 2{s\over u}
+ {t\over u} \Bigr] H(t,u)
- {8\over 3} \pi^2 {s\over t+u}
  \Bigl[ 2{s\over t} + {t\over s} - {u\over s} \Bigr] \\
\noalign{\vskip 5pt}
&+& 4 \Bigl[ 16 - 16{u\over t+u} - 16{s^2\over u(t+u)} - 17{s\over u}
            - {t\over u} + 2{s\over t+u}
            + {s\over s+t} \Bigr] \nonumber \\
\noalign{\vskip 5pt}
&-& 4\ln\Bigl({s\over M_W^2}\Bigr) \Bigl[ 3{t\over u} + 2{s\over u}
+ 4{s\over u}{(t+s)\over (t+u)} + 2{t\over u} {s^2\over (t+u)^2}
    \Bigr] \nonumber \\
\noalign{\vskip 5pt}
&+& 4\ln\Bigl({-u\over M_W^2}\Bigr) \Bigl[ {4s+u\over s+t}
 + {su\over (s+t)^2} \Bigr] \>,
\nonumber
\end{eqnarray}
and
\begin{eqnarray}
H(t,u) = &\phantom{-}& \pi^2
- \ln\Bigl({s\over M_W^2}\Bigr)^2
+ \ln\Bigl({-u\over s}\Bigr)^2
- \ln\Bigl({-u\over M_W^2}\Bigr)^2 \\
\noalign{\vskip 5pt}
&-& 2\, {\rm Li}_2\Bigl(1- {s\over M_W^2}\Bigr)
 -  2\, {\rm Li}_2\Bigl(1- {u\over M_W^2}\Bigr) \>. \nonumber
\end{eqnarray}
The $W$-boson mass is denoted by $M_W^{}$,
$N_C = 3$ is the number of colors, $e$ is the electromagnetic coupling
constant, $U_{q_1 q_2}$ is the Cabibbo-Kobayashi-Maskawa quark
mixing matrix, $x_{\hbox{\scriptsize w}} = \sin^2 \theta_{\hbox{\scriptsize
w}}$ where $\theta_{\hbox{\scriptsize w}}$ is the weak mixing angle,
and $Q_1$ and $Q_2$ are the electric charges of $q_1$ and $q_2$ in units of the
proton charge $e$.
The $2\to 2$ subprocess is labeled by $q_1(p_1) + \bar q_2(p_2) \to
W(p_3) + \gamma(p_4)$ and the parton level kinematic invariants $s,t,u$ are
defined by
\begin{eqnarray}
\noalign{\vskip 5pt}
s = (p_1 + p_2)^2 \>, \qquad t = (p_1 - p_3)^2 \>,
\qquad u = (p_1 - p_4)^2 \>.
\end{eqnarray}
The function $\hbox{\rm Li}_2 (z)$ is the dilogarithm function
\begin{eqnarray}
\noalign{\vskip 5pt}
\hbox{\rm Li}_2 (z) = - \int_0^1 \ln(1-tz) {dt\over t} =
\sum_{k=1}^{\infty} {z^k \over k^2} \>.
\end{eqnarray}

The three-body contribution to the NLO cross section is
\widetext
\FL
\begin{eqnarray}
\noalign{\vskip 5pt}
\sigma^{\phantom{\hbox{\scriptsize NLL}}}_{\hbox{\scriptsize 3 body}}
(p\,p\hskip-7pt\hbox{$^{^{(\!-\!)}}$} &\rightarrow& W \gamma + X) =
\sum_{a,b,c} \int d\hat\sigma(ab \rightarrow W \gamma c)  \\
&\times& \Bigl[ G_{a/p}(x_1,M^2)\,G_{b/p\hskip-7pt\hbox{$^{^{(\!-\!)
}}$}}(x_2,M^2)  + (x_1 \leftrightarrow x_2) \Bigr] dx_1 \,dx_2 \>,
\nonumber
\end{eqnarray}
\narrowtext
where the sum is over all partons contributing to the three
subprocesses $q_1\bar q_2\rightarrow W\gamma g$, $q_1 g \rightarrow
W\gamma q_2$, and $g\bar q_2\rightarrow W\gamma \bar q_1$.
The $2 \to 3$ subprocess is labeled by $p_1 + p_2 \to p_3 + p_4 + p_5$ and the
kinematic invariants $s_{ij}$ and $t_{ij}$ are defined by
$s_{ij} = (p_i + p_j)^2$ and $t_{ij} = (p_i - p_j)^2$.
The integration over three-body phase space and $dx_1 \, dx_2$
is done numerically by standard Monte Carlo techniques.  The
kinematic invariants $s_{ij}$ and $t_{ij}$ are first tested for soft
and  collinear singularities.  If an invariant for a subprocess falls
in a soft  or collinear region of phase space, the contribution from
that subprocess is  not included in the cross section.

Except for the virtual contribution, $d\hat\sigma^{\hbox{\scriptsize
virt}}/dv$ in Eq.~(\ref{EQ:NLO}),
the ${\cal O}(\alpha_s)$ corrections are all proportional to the Born cross
section.  It is easy to incorporate the decay $W \to \ell \nu$ into
those terms which
are proportional to the Born cross section; one simply replaces
$d\hat\sigma^{\hbox{\scriptsize Born}} (q_1 \bar q_2 \to W \gamma)$ with
$d\hat\sigma^{\hbox{\scriptsize Born}}
(q_1 \bar q_2 \to W \gamma \to \ell \nu \gamma)$ in Eq.~(\ref{EQ:NLO}).
It is likewise easy to include the $W$-decay in the NLO bremsstrahlung,
the hard collinear, and the real emission
contributions by making analogous replacements.
When working at the amplitude level, the $W$-decay is trivial to
implement; one simply replaces the $W$-boson polarization vector
$\epsilon_\mu (k)$ with the $W\to \ell \nu$ decay current $J_\mu (k)$ in the
amplitude.  Details of the amplitude level calculations for the Born and real
emission subprocesses can be found in Ref.~\cite{VVJET}.

The only term in which it is more difficult to incorporate the $W$-decay is
the virtual contribution.  Rather than undertake the non-trivial task of
recalculating the virtual correction term for the case of a
leptonically decaying $W$-boson, we have instead opted to use the
virtual correction for a real on-shell
$W$-boson which we subsequently decay ignoring spin correlations.
When spin correlations are ignored, the squared matrix element for $W$-boson
production and decay factorizes into separate production and decay squared
matrix elements when the sum over spins is carried out, {\it i.e.},
\widetext
\FL
\begin{eqnarray}
\noalign{\vskip 5pt}
\sum_{\rm spins}|{\cal M}(q_1 \bar q_2 \to W + X \to \ell \nu + X)|^2
\,&\approx&\, \sum_{\rm spins} |{\cal M}(q_1 \bar q_2 \to W + X )|^2  \\
\noalign{\vskip 5pt}
&&\quad \times\, (4 \pi)^2 \, B(W \to \ell \nu) \, \delta(q_{\ell\nu}^2
- M_W^2) \>,
\nonumber
\end{eqnarray}
\narrowtext
where $B(W \to \ell \nu)$ is the $W \to \ell \nu$ branching ratio and
$q_{\ell\nu}^2$ is the squared $\ell\nu$ invariant mass.

Neglecting spin correlations slightly modifies the shapes of the
angular distributions of the final state leptons.
If no angular cuts ({\it e.g.}, rapidity cuts)
are imposed on the final state leptons, then ignoring spin
correlations does not alter the total cross section. For realistic
rapidity cuts, cross sections are changed by typically 10\% if spin
correlations are neglected.
Since the size of the virtual correction is only about
1\% the size of the Born cross section, the overall effect of
neglecting the spin correlations in the virtual correction is expected
to be negligible compared to the $20\% \sim 30\%$ uncertainty from the parton
distribution functions and the choice of the scale $Q^2$. This will be
demonstrated explicitly in Section~IIID.
(Note that spin correlations are included everywhere in the calculation
except in the virtual contribution.)

\newpage
\subsection{Incorporation of the Decay $W \to \ell \nu$}

The results for the NLO calculation of
$p\,p\hskip-7pt\hbox{$^{^{(\!-\!)}}$} \rightarrow W^{\pm}\gamma + X
\rightarrow \ell^\pm \nu \gamma + X$
can now be summarized. The NLO cross section now consists of three-
and four-body final state contributions:
\widetext
\FL
\begin{eqnarray}
\noalign{\vskip 5pt}
\sigma^{\hbox{\scriptsize NLO}}
 (p\,p\hskip-7pt\hbox{$^{^{(\!-\!)}}$} \to W \gamma + X \to \ell \nu
\gamma + X) = &\phantom{+}&
\sigma^{\hbox{\scriptsize NLO}}_{\hbox{\scriptsize 3 body}}
(p\,p\hskip-7pt\hbox{$^{^{(\!-\!)}}$} \to W \gamma \to \ell \nu
\gamma) \label{EQ:NLOCROSSDECAY} \\
\noalign{\vskip 5pt}
&+& \sigma_{\hbox{\scriptsize 4 body}}
(p\,p\hskip-7pt\hbox{$^{^{(\!-\!)}}$} \to W \gamma + X \to \ell \nu
\gamma + X) \>.
\nonumber
\end{eqnarray}
\narrowtext
The three-body contribution is
\widetext
\FL
\begin{eqnarray}
\noalign{\vskip 5pt}
& &\sigma^{\hbox {\scriptsize NLO}}_{\hbox{\scriptsize 3 body}}
 (p\,p\hskip-7pt\hbox{$^{^{(\!-\!)}}$} \to W \gamma \to \ell \nu \gamma) =
   \sigma^{\hbox{\scriptsize NLO}}_{\hbox{\scriptsize brem}}
 + \sigma^{\hbox{\scriptsize hc}}
 + \sum_{q_1, \bar q_2} \int dv\,dx_1\,dx_2  \label{EQ:TWELVE} \\
\noalign{\vskip 5pt}
 & &\quad \times \biggl[ G_{q_1/p}(x_1,M^2)\,G_{\bar q_2/p\hskip-7pt
\hbox{$^{^{(\!-\!)}}$}}(x_2,M^2) \,
 {d\hat\sigma^{{\hbox {\scriptsize NLO}}}\over dv}
 (q_1\bar q_2\rightarrow W \gamma \to \ell \nu \gamma)
 + (x_1 \leftrightarrow x_2) \biggr] \>,
\nonumber
\end{eqnarray}
\narrowtext
where $\sigma^{\hbox{\scriptsize NLO}}_{\hbox{\scriptsize brem}}$ is the NLO
bremsstrahlung cross section defined in Appendix~A,
$\sigma^{\hbox{\scriptsize hc}}$ is the hard collinear remnant contribution
defined in Appendix~B, the sum is over all contributing quark flavors, and
\FL
\begin{eqnarray}
\noalign{\vskip 5pt}
{d\hat\sigma^{{\hbox {\scriptsize NLO}}}\over dv}
 (q_1\bar q_2 \rightarrow W\gamma \to \ell \nu \gamma) = &\phantom{+}&
 {d\hat\sigma^{\hbox{\scriptsize Born}}\over dv}
 (q_1 \bar q_2 \to W\gamma \to \ell \nu \gamma) \,
 \Biggl[ 1 + C_F {\alpha_s(\mu^2) \over 2 \pi} \biggl\{ 4 \ln(\delta_s)^2
 \nonumber \\
\noalign{\vskip 5pt}
&+& 3 \ln\Bigl({s\over M^2}\Bigr)
+ 4 \ln(\delta_s) \ln\Bigl({s\over M^2}\Bigr)
 \label{EQ:NEWNLO} \\
\noalign{\vskip 5pt}
&+& \lambda_{FC} \Bigl( 9 + {2\over 3} \pi^2 + 3 \ln(\delta_s)
- 2 \ln(\delta_s)^2 \Bigr) \biggr\} \Biggl] \nonumber \\
\noalign{\vskip 5pt}
&+& {d\hat\sigma^{\hbox{\scriptsize virt}}\over dv}
(q_1 \bar q_2 \to W\gamma) \, B(W \to \ell \nu) \>.
\nonumber
\end{eqnarray}
The virtual contribution $d\hat\sigma^{\hbox{\scriptsize virt}} / dv
(q_1 \bar q_2 \to W\gamma)$, which is defined in Eq.~(\ref{EQ:VIRT}),
is multiplied here by the $W \to \ell \nu$ branching ratio.

The four-body contribution is
\widetext
\FL
\begin{eqnarray}
\noalign{\vskip 5pt}
\sigma^{\phantom{\hbox{\scriptsize NLL}}}_{\hbox{\scriptsize 4 body}}
(p\,p\hskip-7pt\hbox{$^{^{(\!-\!)}}$} &\to & W \gamma + X \to \ell \nu
\gamma + X) =
\sum_{a,b,c} \int d\hat\sigma(ab \to W \gamma c \to \ell \nu \gamma c) \\
\noalign{\vskip 5pt}
&\times& \Bigl[ G_{a/p}(x_1,M^2)\,G_{b/p\hskip-7pt\hbox{$^{^{(\!-\!)
}}$}}(x_2,M^2)
 + (x_1 \leftrightarrow x_2) \Bigr] dx_1 \,dx_2 \>,
\nonumber
\end{eqnarray}
\narrowtext
where the sum is over all partons contributing to the three subprocesses
$q_1 \bar q_2 \rightarrow W \gamma g \to \ell \nu \gamma g$,
$q_1 g \rightarrow W \gamma q_2 \to \ell \nu \gamma q_2$, and
$g \bar q_2 \rightarrow W \gamma \bar q_1 \to \ell \nu \gamma \bar q_1$.
The squared matrix elements for the Born subprocess and the real emission
subprocesses were evaluated numerically via helicity amplitude methods as
described in Ref.~\cite{VVJET}.

\subsection{Incorporation of Anomalous $WW\gamma$ Couplings}

The $WW\gamma$ vertex is uniquely determined in the SM by
$\hbox{\rm SU(2)} \bigotimes \hbox{\rm U(1)}$
gauge invariance.  In $W\gamma$ production both the virtual $W$ and the
decaying onshell $W$ couple to essentially massless fermions, which
insures that effectively $\partial_\mu W^\mu=0$. This condition
together with Lorentz invariance, electromagnetic gauge invariance, and $CP$
conservation, allows two free parameters, $\kappa$ and $\lambda$,
in the $WW\gamma$ vertex.
The most general Lorentz and $CP$ invariant vertex compatible with
electromagnetic gauge invariance is described by the effective
Lagrangian~\cite{LAGRANGIAN}
\FL
\begin{eqnarray}
\noalign{\vskip 5pt}
{\cal L}_{WW\gamma} &=& -i \, e \,
\Biggl[ W_{\mu\nu}^{\dagger} W^{\mu} A^{\nu}
              -W_{\mu}^{\dagger} A_{\nu} W^{\mu\nu}
+ \kappa W_{\mu}^{\dagger} W_{\nu} F^{\mu\nu}
+ {\lambda \over M_W^2} W_{\lambda \mu}^{\dagger} W^{\mu}_{\nu} F^{\nu\lambda}
\Biggr] \>,
\label{EQ:LAGRANGE}
\end{eqnarray}
\narrowtext
where $A^{\mu}$ and $W^{\mu}$ are the photon and $W^-$ fields, respectively,
$W_{\mu\nu} = \partial_{\mu}W_{\nu} - \partial_{\nu}W_{\mu}$, and
$F_{\mu\nu} = \partial_{\mu}A_{\nu} - \partial_{\nu}A_{\mu}$.
All higher dimensional operators are obtained by replacing $W^\mu$ with
$(\partial^2)^m W^\mu$, where $m$ is an arbitrary positive integer,
in the terms proportional to $\Delta\kappa=\kappa-1$ and $\lambda$.
These operators form a complete set and can be summed up by
replacing $\Delta\kappa$ and $\lambda$ by momentum dependent form factors. All
details are contained in the specific functional form of the form factor
and its scale $\Lambda$. The form factor nature of $\Delta\kappa$ and
$\lambda$ will be discussed in more detail later.

In Eq.~(\ref{EQ:LAGRANGE}), without loss of generality, we have chosen the
$W$ boson mass $M_W$ as the energy scale in the denominator of the term
proportional to $\lambda$. If a different mass scale, ${\tt M}$, had
been used in Eq.~(\ref{EQ:LAGRANGE}),
then all of our subsequent results could be obtained by scaling
$\lambda$ by a factor ${\tt M}^2/M_W^2$.

The variables $\kappa$ and $\lambda$ are related to the magnetic dipole
moment, $\mu_W^{}$, and the electric quadrupole moment, $Q_W^{}$, of
the $W$-boson:
\begin{eqnarray}
\mu_{W}^{} &=& {e \over 2 M_W^{} }\, (1 + \kappa + \lambda) \>, \\
\noalign{\vskip 5pt}
Q_{W}^{} &=& -{e \over M_W^2 } \, (\kappa - \lambda) \>.
\end{eqnarray}
At tree level in the SM, $\kappa = 1$ and $\lambda = 0$.
The two $CP$ conserving couplings have recently been measured by the UA2
Collaboration in the process $p\bar p\rightarrow e^\pm\nu\gamma X$ at
the CERN $p\bar p$ collider~\cite{UATWO}:
\newcommand{\crc}{\crcr\noalign{\vskip -8pt}}
\begin{eqnarray}
\noalign{\vskip 5pt}
\kappa=1\matrix{+2.6\crc -2.2}~~({\rm for}~\lambda=0) \>, \hskip 1.cm
\lambda=0\matrix{+1.7\crc -1.8}~~({\rm for}~\kappa=1) \>,
\end{eqnarray}
at the 68.3\% confidence level (CL). Although bounds on these couplings
can also be extracted from
low energy data and high precision measurements at the $Z$ pole, there are
ambiguities and model dependencies in the results \cite{DE,BL,HISZ}.
{}From loop contributions
to $(g-2)_\mu$ one estimates~\cite{MUON} limits which are typically
of ${\cal O}(1-10)$. No rigorous bounds on $WW\gamma$ couplings can be
obtained from LEP~I data if correlations between different
contributions to the anomalous couplings are fully taken into account.
Without serious cancelations among various one loop contributions, one
finds~\cite{HISZ,CRS} $|\Delta\kappa|,~|\lambda|\leq 0.5-1.5$
at the 90\% CL from present data on $S$, $T$, and $U$~\cite{PT} (or,
equivalently, $\epsilon_1$, $\epsilon_2$, and $\epsilon_3$~\cite{ALT}).
In contrast, one expects deviations from the SM of ${\cal O}(10^{-2})$
or less for $\kappa$ and $\lambda$ if an approach based on chiral perturbation
theory~\cite{BDV} is used.

If $CP$ violating $WW\gamma$ couplings are allowed, two additional
free parameters, $\tilde\kappa$ and $\tilde\lambda$ appear in the
effective Lagrangian. However,
$CP$ violating operators are tightly constrained by measurements of the neutron
electric dipole moment which restrict $\tilde \kappa$ and $\tilde \lambda$ to
$|\tilde \kappa|$, $|\tilde \lambda| < {\cal
O}(10^{-3})$~\cite{CPLIMITS}. $CP$ violating $WW\gamma$ couplings are,
therefore, not considered in this paper.

The Feynman rule for the $WW\gamma$ vertex factor
corresponding to the Lagrangian in Eq.~(\ref{EQ:LAGRANGE}) is
\FL
\begin{eqnarray}
-i\, e \, (Q_1-Q_2)\, \Gamma_{\beta \mu \nu}^{} (k, k_1, k_2) =
&-&i\, e \, (Q_1-Q_2) \label{EQ:NSMCOUPLINGS} \\
\noalign{\vskip 5pt}
&\times&  \biggl[
\Gamma_{\beta \mu \nu}^{\hbox{\scriptsize SM}} (k, k_1, k_2)
+ \Gamma_{\beta \mu \nu}^{\hbox{\scriptsize NSM}} (k, k_1, k_2) \biggr] \>,
\nonumber
\end{eqnarray}
where the labeling conventions for the four-momenta and Lorentz
indices are defined by Fig.~\ref{FIG:VERTEX},
$(Q_1 - Q_2)$ is the electric charge of the $W$-boson
($Q_1$ and $Q_2$ are the electric charges of $q_1$ and $q_2$ in units
of the proton charge $e$), and the factors
$\Gamma^{\hbox{\scriptsize SM}}$ and $\Gamma^{\hbox{\scriptsize NSM}}$
are the SM and non-standard model vertex factors:
\FL
\begin{eqnarray}
\noalign{\vskip 5pt}
\Gamma_{\beta \mu \nu}^{\hbox{\scriptsize SM}} (k, k_1, k_2) =
&\phantom{+}& (k_1 - k_2)_{\beta} \, g_{\nu \mu} + 2 \,
 k_{\mu} \, g_{\beta \nu} - 2 \, k_{\nu} \, g_{\beta \mu} \>, \\
\noalign{\vskip 5pt}
\Gamma_{\beta \mu \nu}^{\hbox{\scriptsize NSM}} (k, k_1, k_2)
= &\phantom{+}&  {1\over 2} \,
\left( \Delta\kappa + \lambda {k^2 \over M_W^2} \right)
        (k_1 - k_2)_{\beta} \, g_{\nu \mu} \\
\noalign{\vskip 5pt}
&-& {\lambda \over M_W^2} \, (k_1 - k_2)_{\beta} \, k_{\nu} \, k_{\mu}
 + (\Delta\kappa + \lambda) \, k_{\mu} \, g_{\beta \nu} \>.
\nonumber
\end{eqnarray}
\narrowtext
The non-standard model vertex factor is written here in terms of
$\Delta \kappa = \kappa - 1$ and $\lambda$, which both vanish in the SM.

It is straight forward to include the non-standard model couplings in
the amplitude level calculations.  Using the computer algebra program
FORM~\cite{FORM}, we have computed the $q_1 \bar q_2 \to W \gamma$
virtual correction with the modified vertex
factor of Eq.~(\ref{EQ:NSMCOUPLINGS}), however,
the resulting expression is too lengthy to present here. The
non-standard $WW\gamma$ couplings of Eq.~(\ref{EQ:LAGRANGE}) do not
destroy the renormalizability of QCD. Thus, the infrared singularities
from the soft and virtual contributions are explicitly canceled, and the
collinear singularities are factorized and absorbed into the definition
of the parton distribution and photon fragmentation functions, exactly
as in the SM case.

The anomalous couplings can not be simply inserted into the vertex
factor as constants because this would violate $S$-matrix unitarity.
Tree level unitarity uniquely restricts the $WW\gamma$ couplings to their
SM gauge theory values at asymptotically high energies~\cite{CORNWALL}.
This implies that any deviation of $\Delta \kappa$ or $\lambda$ from the SM
expectation has to be described by a form factor
$\Delta \kappa(M_{W\gamma}^2, p_W^2, p_\gamma^2)$ or
$\lambda(M_{W\gamma}^2, p_W^2, p_\gamma^2)$
which vanishes when either the square of the $W\gamma$ invariant mass,
$M_{W\gamma}^2$, or the
square of the four-momentum of the final state $W$ or photon ($p_W^2$ or
$p_\gamma^2)$ becomes large.  In $W\gamma$ production $p_\gamma^2 = 0$ and
$p_W^2 \approx M_W^2$ even when the finite $W$-width is taken into account.
However, large values of $M_{W\gamma}^2$ will be probed at future hadron
colliders like the LHC or the SSC and the $M_{W\gamma}^2$ dependence
of the anomalous couplings has
to be included in order to avoid unphysical results which would violate
unitarity.  Consequently, the anomalous couplings are introduced
via form factors~\cite{BAURZEP,WILLEN}
\begin{eqnarray}
\noalign{\vskip 5pt}
\Delta \kappa(M_{W\gamma}^2, p_W^2 = M_W^2, p_\gamma^2 = 0) \> &=& \>
{\Delta \kappa_0 \over (1 + M_{W\gamma}^2/\Lambda^2)^n } \>,
\label{EQ:KAPPAFORM} \\
\noalign{\vskip 5pt}
\lambda(M_{W\gamma}^2, p_W^2 = M_W^2, p_\gamma^2 = 0) \> &=& \>
{\lambda_0 \over (1 + M_{W\gamma}^2/\Lambda^2)^n } \>,
\label{EQ:LAMBDAFORM}
\end{eqnarray}
where $\Delta \kappa_0$ and $\lambda_0$ are the form factor values at
low energies and
$\Lambda$ represents the scale at which new physics becomes important in the
weak boson sector, {\it e.g.}, due to a composite structure of the $W$-boson.
In order to guarantee unitarity, $n>1/2$ for $\Delta\kappa$ and $n>1$
for $\lambda$. For the numerical results presented here, we use a dipole
form factor ($n=2$) with a scale $\Lambda = 1$~TeV. The exponent $n=2$
is chosen in order to suppress $W\gamma$ production at energies
$\sqrt{\hat s}\gg\Lambda\gg M_W$, where novel phenomena like resonance
or multiple weak boson production are expected to become important.

\section{PHENOMENOLOGICAL RESULTS}

We shall now discuss the phenomenological implications of NLO QCD
corrections to $W\gamma$ production at the Tevatron
($p\bar p$ collisions at $\sqrt{s} = 1.8$~TeV) and the SSC ($pp$
collisions at $\sqrt{s} = 40$~TeV). We first briefly describe
the input parameters, cuts, and the finite energy resolution smearing
used to simulate detector response. We then discuss in detail the impact
of NLO QCD corrections on the observability of non-standard $WW\gamma$
couplings in $W\gamma$ production at the Tevatron and SSC. To simplify
the discussion, we shall concentrate on $W^+\gamma$ production. In
$p\bar p$ collisions the rates for $W^+\gamma$ and $W^-\gamma$
production are equal. At $pp$ colliders, the $W^-\gamma$ cross section
is slightly smaller than that of $W^+\gamma$ production. Furthermore,
we shall only consider $W\to e\nu$ decays in the following. Since results
and conclusions for $W\gamma$ production at the LHC are qualitatively
very similar to those obtained for the SSC, we do not show differential
distributions for LHC energies.

\subsection{Input Parameters}

The numerical results presented in this
section were obtained using the two-loop expression for
$\alpha_s$. The QCD scale $\Lambda_{\hbox{\scriptsize QCD}}$
is specified for four
flavors of quarks by the choice of parton distribution functions  and
is adjusted whenever a heavy quark threshold is crossed so that
$\alpha_s$ is a continuous function of $Q^2$. The heavy quark masses
were taken to be $m_b=5$~GeV and $m_t=150$~GeV.
The SM parameters used in our numerical simulations are $M_Z = 91.173$~GeV,
$M_W = 80.22$~GeV, $\alpha (M_W) =1/128$, and $\sin^2
\theta_{\hbox{\scriptsize w}} = 1 - (M_W^{}/M_Z^{})^2$. These values are
consistent with recent measurements at LEP, the CERN $p\bar p$
collider, and the Tevatron \cite{LEP,MT,MW}. The soft and collinear
cutoff parameters are fixed to $\delta_s = 10^{-2}$ and $\delta_c =
10^{-3}$ unless stated otherwise. The parton
subprocesses have been summed over $u,d,s$, and $c$ quarks and the
Cabibbo mixing angle has been chosen such that $\cos^2 \theta_C =
0.95$. The leptonic branching ratio has been taken to be $B(W \to e
\nu) = 0.109$ and the total width of the $W$-boson is $\Gamma_W =
2.12$~GeV. Except where otherwise stated, a single scale
$Q^2=M^2_{W\gamma}$,  where $M_{W\gamma}$ is the invariant mass of the
$W\gamma$ pair, has been used for the renormalization scale $\mu^2$
and the factorization scale $M^2$.

In order to get consistent NLO results it is necessary to use parton
distribution functions which have been fit to next-to-leading order. In
our numerical simulations we have used
the Martin-Roberts-Stirling (MRS)~\cite{MRS} set S0 distributions
with $\Lambda_4 = 215$~MeV, which take into account the most recent
NMC~\cite{NMC} and CCFR~\cite{CCFR} data. The MRS
distributions are defined in the universal (${\rm \overline{MS}}$)
scheme and thus the factorization defining parameter $\lambda_{FC}$ in
Eqs.~(\ref{EQ:NLO}), (\ref{EQ:NEWNLO}), and (\ref{EQ:AP})
should be $\lambda_{FC} = 0$. For convenience,
the MRS set S0 distributions have also been used for the LO calculations.

\subsection{Cuts}

The cuts imposed in our numerical simulations are motivated by two factors:
1)~the finite acceptance and resolution of the detector and
2)~the need to suppress radiative $W$ decay which results in the same
final state as $W\gamma$ production. The finite
acceptance of the detector is simulated by cuts on the four-vectors of the
final state particles.  This group of cuts includes requirements on the
transverse momentum of the photon and electron, and on the missing
transverse momentum, $p\llap/_T^{}$, associated with the neutrino.
Also included in this group are cuts on the pseudorapidity, $\eta$, of the
photon and electron. In addition, the electron and photon are
also required to be separated in the pseudorapidity-azimuthal-angle plane
\begin{eqnarray}
\noalign{\vskip 5 pt}
\Delta R (e,\gamma) = \left[ \left(\Delta \phi_{e \gamma}^{} \right)^2
	                      + \left(\Delta \eta_{e \gamma}^{} \right)^2
\right]^{1/2} \>.
\end{eqnarray}

Since we ignore photon radiation from the final state lepton line in our
calculation, it is necessary to impose cuts which will efficiently suppress
contributions from this diagram.
In radiative $W$ decays the lepton photon separation sharply peaks at
small values due to the collinear singularity associated with the
diagram in which the photon is radiated from the final state lepton
line.  In the following we shall therefore impose a large separation cut
of $\Delta R(e,\gamma)>0.7$. Contributions from $W\to e\nu\gamma$
can be further reduced by a cluster transverse mass cut.
In radiative $W$ decays the $e \nu$ pair and the photon form a
system with invariant mass $M(e \nu \gamma)$ close to $M_W^{}$, whereas for
$W\gamma$ production $M(e \nu \gamma)$ is always larger than $M_W^{}$ if
finite $W$-width effects are ignored.
This difference suggests that an $M(e \nu \gamma)$ cut can be used to
separate $e \gamma p\llap/_T^{}$ events originating from radiative $W$
decays from $e \gamma p\llap/_T^{}$ events originating from
$W\gamma$ events. However, because of the nonobservation of the neutrino,
$M(e \nu \gamma)$ cannot be determined unambiguously and the minimum
invariant mass or the cluster transverse mass~\cite{CTMASS} is more useful:
\FL
\begin{eqnarray}
M_T^2 (e \gamma;p\llap/_T^{}) =
\biggl[ \Bigl( M_{e \gamma}^2
 + \bigl| \hbox{\bf p}_{T}^{}(\gamma)
 + \hbox{\bf p}_{T}^{}(e) \bigr|^2 \Bigr)^{1/2}
 + p\llap/_T^{} \biggr]^2
- \bigl|   \hbox{\bf p}_{T}^{}(\gamma)
         + \hbox{\bf p}_{T}^{}(e)
         + \hbox{\bf p}\llap/_T^{} \bigr|^2 \, . \> \>
\end{eqnarray}
Here $M_{e \gamma}$ denotes the invariant mass of the $e \gamma$ pair.
For $W \rightarrow e \nu \gamma$ the cluster transverse mass peaks sharply
at $M_W^{}$ (Ref.~\cite{CTMASS}) and drops rapidly above the $W$ mass.
Thus $e \gamma p\llap/_T^{}$ events originating from $W\gamma$ production and
radiative $W$ decays can be distinguished if $M_T^{}(e \gamma;p\llap/_T^{})$
is cut slightly above $M_W^{}$ (Ref.~\cite{CHH}).  In our numerical results we
thus require
\begin{eqnarray}
M_T^{}(e \gamma;p\llap/_T^{}) > 90\ \hbox{\rm GeV} \>.
\end{eqnarray}
As shown in Ref.~\cite{BAURBERGER}, this cut, together with the lepton
photon separation cut, is quite efficient in suppressing
radiative $W$ decay events.

At leading order, $W\gamma$ events are produced not only by the Born subprocess
$q_1 \bar q_2 \rightarrow W \gamma$ but also by the photon bremsstrahlung
process which proceeds via subprocesses such as $q_1 g \rightarrow W q_2$
followed by photon bremsstrahlung from the final state quark.
As demonstrated in Ref.~\cite{BREM}, the bremsstrahlung process is not only
significant, but is in fact the dominant production mechanism at supercollider
center of mass energies.  However, the bremsstrahlung process does not involve
the $WW\gamma$ vertex and is thus a background to the Born process which
is sensitive to the $WW\gamma$ coupling.
Fortunately, the photon bremsstrahlung events can be suppressed by requiring
the photon to be isolated~\cite{BREM}. A photon isolation cut typically
requires the sum of the hadronic energy $E_{\hbox{\scriptsize had}}$ in a
cone of size $R_0$ about the direction of the photon to be less than a
fraction $\epsilon_{\hbox{\scriptsize h}}$ of
the photon energy $E_{\gamma}$, {\it i.e.},
\begin{eqnarray}
\sum_{\Delta R < R_0} \, E_{\hbox{\scriptsize had}} <
\epsilon_{\hbox{\scriptsize h}} \, E_{\gamma} \>,
\label{EQ:ISOL}
\end{eqnarray}
with $\Delta R = [(\Delta \phi)^2 + (\Delta \eta)^2 ]^{1/2}$.
To suppress the photon bremsstrahlung background, a photon isolation cut
with $\epsilon_{\hbox{\scriptsize h}}=0.15$~\cite{EPS} will
be applied in the numerical results presented in this section. For this
value of $\epsilon_{\hbox{\scriptsize h}}$, the photon bremsstrahlung
background is less than 10\% of the Born $W\gamma$ signal rate.

The complete set of cuts can now be summarized as follows.
\begin{quasitable}
\begin{tabular}{cc}
Tevatron & SSC\\
\tableline
$p_{T}^{}(\gamma)       > 10$~GeV  & $p_{T}^{}(\gamma)       > 100$~GeV\\
$p_{T}^{}(e)         > 20$~GeV  & $p_{T}^{}(e)         > 25$~GeV\\
$p\llap/_T^{}           > 20$~GeV  & $p\llap/_T^{}           > 50$~GeV\\
$|\eta(\gamma)|         < 1.0$     & $|\eta(\gamma)|         < 2.5$\\
$|\eta(e)|           < 2.5$     & $|\eta(e)|           < 3.0$\\
$\Delta R (e,\gamma) > 0.7$     & $\Delta R (e,\gamma) > 0.7$\\
$M_T^{}(e \gamma;p\llap/_T^{}) > 90$~GeV &
$M_T^{}(e \gamma;p\llap/_T^{}) > 90$~GeV \\
${\sum \atop \Delta R < 0.7} \, E_{\hbox{\scriptsize h}}
 < 0.15 \, E_{\gamma}$ &
${\sum \atop \Delta R < 0.7} \, E_{\hbox{\scriptsize h}}
 < 0.15 \, E_{\gamma}$
\end{tabular}
\end{quasitable}

The effects of non-standard $WW\gamma$ couplings are most pronounced in
the central photon rapidity region. We therefore impose a rather
stringent cut on $\eta(\gamma)$, in particular at the Tevatron. The large
$p_T^{}(\gamma)$ and $p\llap/_T^{}$ cuts at SSC energies are chosen to
reduce potentially dangerous backgrounds from
$W+1$~jet production, where the jet is misidentified as a photon, and
from processes where particles outside the rapidity range covered by the
detector contribute to the missing transverse momentum. Present
studies~\cite{FAKE,RSDC} indicate that these backgrounds are under control for
$p_T^{}(\gamma) > 100$~GeV and $p\llap/_T^{}>50$~GeV.

\subsection{Finite Energy Resolution Effects}

Uncertainties in the energy measurements of the charged lepton and the
photon in the detector are simulated by Gaussian smearing of the particle
four-momentum vector with standard deviation $\sigma$ in our
calculation.  For distributions which
require a jet definition, {\it e.g.}, the $W\gamma + 1$~jet exclusive cross
section, the jet four-momentum vector is also smeared.  The standard
deviation $\sigma$
depends on the particle type and the detector.  The numerical results
presented here for the Tevatron and SSC center of mass energies
were made using $\sigma$ values based on the CDF and SDC
specifications, respectively~\cite{RSDC,RCDF}.

\subsection{Inclusive NLO Cross Sections}

The sensitivity of $W\gamma$ production to anomalous $WW\gamma$
couplings in the Born approximation was studied in detail in
Refs.~\cite{BAURZEP} and~\cite{BAURBERGER}. The
photon transverse momentum distribution, $d\sigma/dp_{T}^{}(\gamma)$,
the photon rapidity spectrum in the parton center of mass frame,
$d\sigma/d|y^*_{\gamma}|$, and the $W\gamma$ invariant mass
differential cross section, $d\sigma/dM_{W\gamma}^{}$, were found to be
sensitive to the anomalous couplings.
Of these three distributions, the $p_{T}^{}(\gamma)$ distribution is the most
sensitive indicator of anomalous couplings since it is a directly observable
quantity.  On the other hand, the $y^*_{\gamma}$ and
$M_{W\gamma}^{}$ distributions
can only be reconstructed with a two-fold ambiguity corresponding to the two
solutions for the longitudinal momentum of the neutrino.  Thus the sensitivity
of these two distributions to anomalous couplings is degraded.

At hadron colliders the $W\gamma$ invariant mass cannot be determined
unambiguously because the neutrino from the $W$ decay is not observed.
If the transverse momentum of the neutrino is identified with the missing
transverse momentum of a given $W\gamma$ event, the unobserved longitudinal
neutrino momentum $p_L^{}(\nu)$ can be reconstructed, albeit with a twofold
ambiguity, by imposing the constraint that the neutrino and the charged lepton
four-momenta combine to form the $W$ rest mass~\cite{CHH,STROUGHAIR}.
Neglecting the electron mass one finds
\begin{eqnarray}
\noalign{\vskip 5pt}
p_L^{}(\nu) &=& {1\over 2\, p^2_T(e) } \Biggl\{ p_L^{}(e)
\Bigl(M_W^2
+ 2\, \hbox{\bf p}_T^{}(e) \cdot \hbox{\bf p}\llap/_T^{} \Bigr)  \\
\noalign{\vskip 5pt}
&&\qquad\qquad \pm \, p(e)\, \biggl[ \Bigl(M_W^2
+ 2\, \hbox{\bf p}_T^{}(e) \cdot
\hbox{\bf p}\llap/_T^{} \Bigr)^2
- 4\, p_T^2(e)\, p\llap/_T^2 \biggr]^{1/2} \Biggr\}
\>,
\nonumber
\end{eqnarray}
where $p_L(e)$ denotes the longitudinal momentum of the electron.
The two solutions for $p_L^{}(\nu)$ are used to reconstruct
two values for $M_{W\gamma}^{}$ and $y^*_{\gamma}$. Both values are then
histogrammed, each with half the event weight.

The dependence of the total cross section on the collinear and soft
cutoff parameters is illustrated in Fig.~\ref{FIG:DELTASC} which shows
the total NLO cross section for $pp \to W^+ \gamma + X \to e^+ \nu
\gamma + X$ plotted versus $\delta_c$ and $\delta_s$, for $\sqrt{s} =
40$~TeV and the cuts described in Section~IIIB. The $n$- and $n+1$-body
contributions are also plotted for illustration ($n=3$ for this process).
The figure shows that the $3$- and $4$-body
contributions, which separately have no physical meaning, vary strongly with
$\delta_c$ and $\delta_s$, however, the total cross section, which is the sum
of the $3$- and $4$-body contributions, is independent of $\delta_c$
and $\delta_s$ over a wide range of these parameters.

The differential cross section for $p_T^{} (\gamma)$ in the reaction
$p \, \bar p \rightarrow W^{+} \gamma + X \rightarrow e^+ \nu_e \gamma + X$
at $\sqrt{s} = 1.8$~TeV
is shown in Fig.~\ref{FIG:PTGTEV}.  The Born and NLO results are shown in
Fig.~\ref{FIG:PTGTEV}a and Fig.~\ref{FIG:PTGTEV}b, respectively.  In
both cases, results are displayed for the SM and for
two sets of anomalous couplings, namely, $(\lambda_0 = 0.5,
\Delta\kappa_0 = 0)$ and $(\lambda_0 = 0, \Delta\kappa_0 = 1.0)$.  For
simplicity, only one anomalous
coupling at a time is allowed to differ from its SM value.
The figure shows that at the Tevatron center of mass energy,
NLO QCD corrections do not have a large influence on the sensitivity of
the photon transverse momentum distribution to anomalous couplings.
Closer inspection reveals, however, that the shape of
$d\sigma/dp_T^{}(\gamma)$ is changed somewhat in the SM case,
while it remains essentially unmodified for non-standard couplings.
The ${\cal O}(\alpha_s)$ corrections at Tevatron energies are
approximately 30\% for the SM as well as for the
anomalous coupling cases at small photon transverse momenta. In the
SM case, the size of the QCD corrections increases to $\sim
60\%$ at large values of $p_T^{}(\gamma)$, whereas they stay essentially at
the 30\% level for (sufficiently large) non-standard $WW\gamma$ couplings.
Since the anomalous terms in the helicity amplitudes grow like
$\sqrt{\hat s}/M_W$ ($\hat s/M_W^2$) for $\Delta\kappa$ ($\lambda$),
non-standard couplings give large enhancements in the cross
section at large values of $p_T^{} (\gamma)$.

Figure~\ref{FIG:MWGTEV} shows the reconstructed invariant mass
distribution of the
$W\gamma$ system for the same set of parameters as in the previous
figure. The Born and NLO cross sections
again display similar sensitivity to the effects of anomalous couplings.
The shape change of the SM invariant mass distribution is
less pronounced than in $d\sigma/dp_T^{}(\gamma)$.

The size of the ${\cal O}(\alpha_s)$ QCD corrections becomes more obvious
in the photon rapidity distribution in the reconstructed parton center
of mass frame, $d\sigma/d|y^*_\gamma|$, which is shown
in Fig.~\ref{FIG:YGTEV}. The parameters are again the same as in
Fig.~\ref{FIG:PTGTEV}. The pronounced dip at $|y^*_{\gamma}| = 0$ in
the SM case can be understood as a consequence of the radiation
amplitude zero (RAZ). For $u\bar d\to W^+\gamma$ ($d\bar u\to
W^-\gamma$) all contributing
helicity amplitudes vanish for $\cos\Theta=-1/3$ (+1/3), where
$\Theta$ is the angle between the quark and the photon in the parton
center of mass frame. As a result, $d\sigma/d|y^*_\gamma|$, develops a
dip at $|y^*_\gamma|=0$. The inclusion of anomalous couplings at the $WW\gamma$
vertex destroys the RAZ and the dip is, at least partially, filled.
Comparison of Figs.~\ref{FIG:YGTEV}a and~\ref{FIG:YGTEV}b shows that NLO QCD
corrections and anomalous $WW\gamma$ couplings affect the
$|y^*_\gamma|$ distribution in a qualitatively similar way.
Next-to-leading log QCD corrections, however, do not completely obscure
the dip at $|y^*_\gamma|=0$. At Tevatron energies, the dominant
contribution to the NLO cross section originates from quark-antiquark
annihilation. Apart from the photon bremsstrahlung contribution, $\sigma^{\rm
NLO}_{\rm brem}$ [see Eq.~(\ref{EQ:TWELVE})], which is strongly
suppressed by the photon isolation cut, Eq.~(\ref{EQ:ISOL}),
all $2\rightarrow 2$ terms
are proportional to the $q_1\bar q_2\rightarrow W\gamma$ matrix element
in the Born approximation and, therefore, preserve the radiation zero.
Furthermore, the $2\rightarrow 3$ process $q_1\bar q_2\rightarrow
W^\pm\gamma g$ exhibits a RAZ at $\cos\Theta=\mp 1/3$ if the gluon is
collinear with the photon~\cite{BRO}, and also in the soft gluon limit,
$E_g\rightarrow 0$.

The $p_T^{}(\gamma)$ differential cross section, the reconstructed
$W\gamma$ invariant mass
distribution, and the $|y_\gamma^*|$ distribution for $W^+\gamma$
production at the SSC are shown in Figs.~\ref{FIG:PTGSSC}
--~\ref{FIG:YGSSC}. Qualitatively similar results are also obtained
for $W^-\gamma$ production.
Results are shown for the SM (solid line) and for two sets of anomalous
couplings, namely, $(\lambda_0 = 0.25, \Delta\kappa_0 = 0)$ (dashed
line) and $(\lambda_0 = 0, \Delta\kappa_0 = 1.0)$ (dotted line). Due to
the form factor parameters assumed,
the result for $\Delta\kappa_0=1$ approaches the SM result at
large values of $p_T^{}(\gamma)$ and $M_{W\gamma}$. As mentioned before, we
have used $n=2$ and a form factor scale of $\Lambda=1$~TeV in all our
numerical simulations [see Eqs.~(\ref{EQ:KAPPAFORM})
and~(\ref{EQ:LAMBDAFORM})]. For a larger scale $\Lambda$, the
deviations from the SM result become more pronounced at high energies
and transverse momenta (see Ref.~\cite{BAURZEP} for details).

At SSC energies, the inclusive
NLO QCD corrections are very large, most notably in the SM case. The shape
of the $p_T^{}(\gamma)$ distribution is significantly affected by the ${\cal
O}(\alpha_s)$ corrections. For
$p_T^{}(\gamma)=1$~TeV, the QCD corrections increase the SM cross section by
more than one order of magnitude. In the presence of anomalous
couplings, the higher order QCD corrections are smaller than in the
SM, although they are still large. Thus, at next-to-leading order, the
sensitivity of the photon transverse momentum spectrum
to anomalous couplings is severely reduced;
the same is true, although to a smaller degree, for the
$W\gamma$ invariant mass distribution (see Fig.~\ref{FIG:MWGSSC}).
The low invariant mass tail in the NLO $M_{W\gamma}$
distribution is due to events where the $W$ boson and the photon are
almost collinear. The dip at $|y^*_\gamma|=0$, indicating
the radiation zero, is completely filled by the QCD
corrections (see Fig.~\ref{FIG:YGSSC}). Note that the $|y^*_\gamma|$
distributions for the NLO SM and the $\Delta\kappa_0=1$
Born approximation are quite similar.

\subsection{Exclusive NLO QCD Corrections and Jet Veto}

The size of the ${\cal O}(\alpha_s)$ QCD corrections at supercollider
energies and their effect on
the shape of the $p_T^{}(\gamma)$ distribution can be understood by
considering the Born process $q_1\bar q_2\to W\gamma$ and the quark
gluon fusion process $q_1g\rightarrow W\gamma q_2$ in
more detail. In the SM, delicate cancelations between the amplitudes of
the three Born diagrams shown in Fig.~\ref{FIG:BORNGRAPHS} occur in the
central rapidity region. These
cancelations are responsible for the radiation zero and suppress the
$W\gamma$ differential cross section, in particular for large photon
transverse momenta.

In the limit $p_T^{}(\gamma)\gg M_W$, the cross section for
$q_1g\rightarrow W\gamma q_2$ can be obtained using the Altarelli-Parisi
approximation for collinear emission. One finds:
\begin{eqnarray}
\noalign{\vskip 5pt}
d\hat\sigma(q_1g\rightarrow W\gamma q_2)=d\hat\sigma(q_1g\rightarrow
q_1\gamma)\,
{g^2_W\over 16\pi^2}\,\ln^2\!\left({p_T^2(\gamma)\over M_W^2}\right)\, ,
\label{EQ:COLLAPPROX}
\end{eqnarray}
where $g_W^{}=e/\sin\theta_{\rm w}$.
Thus, the quark gluon fusion process carries an enhancement factor
$\ln^2(p_T^2(\gamma)/M_W^2)$ at large photon transverse momentum.
It arises from the kinematical region where the photon is
produced at large $p_T^{}$ and recoils against the quark, which radiates a
soft $W$ boson which is almost collinear to the quark. Since the Feynman
diagrams entering the derivation of Eq.~(\ref{EQ:COLLAPPROX})
do not involve the $WW\gamma$ vertex, the logarithmic enhancement
factor only affects the SM matrix elements. At the SSC, the
$p_T^{}(\gamma)$ differential cross section obtained
using Eq.~(\ref{EQ:COLLAPPROX}) agrees within 40\% with the exact photon
transverse momentum distribution for $p_T^{}(\gamma) >300$~GeV.
Together with the very large $qg$ luminosity at supercollider
energies and the suppression of the SM $W\gamma$ rate at large photon
transverse momenta in the Born approximation, the logarithmic
enhancement factor is responsible for the size of the inclusive NLO QCD
corrections to $W\gamma$ production, as well as for the change in the
shape of the $p_T^{}(\gamma)$ distribution. The same enhancement factor
also appears in the antiquark gluon fusion process, however, the $\bar
qg$ luminosity is much smaller than the $qg$ luminosity for large photon
transverse momenta. Since the $W$ does not couple directly to the gluon,
the process $q_1\bar q_2\to W\gamma g$ is not enhanced at large photon
transverse momenta.

{}From the picture outlined in the previous paragraph,
one expects that, to next-to-leading
order at supercollider energies, $W\gamma$ events with a high $p_T^{}$
photon most of the time also contain a high transverse momentum jet. At
the Tevatron, on the other hand, the fraction of high $p_T^{}(\gamma)$
$W\gamma$ events with a hard jet should be considerably smaller, due to
the much reduced $qg$ luminosity at lower energies.
For a given jet definition it is straightforward to split the
inclusive NLO $W\gamma+X$ cross section into the
NLO $W\gamma+0$~jet and the leading order (LO) $W\gamma+1$~jet cross
sections. The decomposition of the inclusive SM NLO $p_T^{}(\gamma)$ and
$|y^*_\gamma|$ differential cross sections into NLO
0-jet and LO 1-jet exclusive cross sections at the Tevatron (SSC) are
shown in Figs.~\ref{FIG:PTGTEVEX}a and~\ref{FIG:YGTEVEX}a
(Figs.~\ref{FIG:PTGSSCEX}a and~\ref{FIG:YGSSCEX}a), respectively.
The SM NLO 0-jet $p_T^{}(\gamma)$ and $|y^*_\gamma|$ distributions at the
Tevatron (SSC) are compared with
the corresponding distributions obtained in the Born approximation in
Figs.~\ref{FIG:PTGTEVEX}b and~\ref{FIG:YGTEVEX}b
(Figs.~\ref{FIG:PTGSSCEX}b and~\ref{FIG:YGSSCEX}b). Here, a jet is
defined as a quark or gluon with
\begin{eqnarray}
p_T^{}(j)>10~{\rm GeV}\hskip 1.cm {\rm and} \hskip 1.cm |\eta(j)|<2.5
\label{EQ:TEVJET}
\end{eqnarray}
at the Tevatron, and
\begin{eqnarray}
p_T^{}(j)>50~{\rm GeV}\hskip 1.cm {\rm and} \hskip 1.cm |\eta(j)|<3
\label{EQ:SSCJET}
\end{eqnarray}
at the SSC. The sum of the NLO 0-jet and the LO 1-jet exclusive cross
section is equal to the inclusive NLO cross section.

With the jet definition of Eq.~(\ref{EQ:TEVJET}), the
inclusive NLO cross section at the Tevatron is composed predominately
of 0-jet events at low $p_T^{}(\gamma)$ (see Fig.~\ref{FIG:PTGTEVEX}a).
Due to the logarithmic enhancement factor, the
1-jet cross section becomes relatively more important at large photon
transverse momenta. For $p_T^{}(\gamma)$ values above 100~GeV the 0-jet
and 1-jet cross sections contribute nearly equally to the
inclusive NLO cross section.
Fig.~\ref{FIG:PTGTEVEX}b compares the NLO $W\gamma+0$~jet cross section
with the result obtained in the Born approximation.
The NLO and Born cross sections are almost equal at small
$p_T^{}(\gamma)$ for the jet definition used here. For large photon
transverse momenta the NLO 0-jet result is about 20\% smaller than the
cross section in the Born approximation. It is obvious from
Fig.~\ref{FIG:PTGTEVEX} that the QCD corrections to the NLO 0-jet
$p_T^{}(\gamma)$ distribution are much smaller than the
inclusive ${\cal O}(\alpha_s)$ corrections.

The results shown in
Fig.~\ref{FIG:PTGTEVEX} were obtained for $Q^2=M^2_{W\gamma}$. Since the
$W\gamma+1$~jet and the $W\gamma+0$~jet cross section in the Born
approximation are tree level results, the shape and
the absolute normalization of the $p_T^{}(\gamma)$ distributions
are sensitive to the choice of the factorization scale
$Q^2$. For $Q^2=M^2_W$, for example, the $W\gamma+1$~jet cross section
is larger than the NLO $W\gamma+0$~jet result for $p_T^{}(\gamma)>70$~GeV.
The $p_T^{}(\gamma)$ differential cross section in the Born approximation
also changes its shape quite considerably. Whereas the result for
$d\sigma/dp_T^{}(\gamma)$ changes very little at small $p_T^{}(\gamma)$, the
differential cross section at $p_T^{}(\gamma)=200$~GeV for $Q^2=M_W^2$ is
about a factor~2 larger than the result for $Q^2=M_{W\gamma}^2$. On the
other hand, the NLO $W\gamma+0$~jet
photon $p_T^{}$ differential cross section is very insensitive to the value
of $Q^2$ chosen. The $Q^2$ dependence of the $W\gamma$ cross section
will be discussed in more detail later.

Figure~\ref{FIG:YGTEVEX}a displays the inclusive NLO, the ${\cal
O}(\alpha_s)$ 0-jet, and the LO
1-jet $|y^*_{\gamma}|$ distributions for Tevatron energies. As we have
observed earlier, the inclusive NLO QCD corrections partially fill in
the dip at $|y^*_\gamma|=0$ which signals the SM radiation
zero. It is clear from Fig.~\ref{FIG:YGTEVEX}a that events with a high
$p_T^{}$ jet are responsible for this effect. The exclusive NLO 0-jet and Born
$|y^*_\gamma|$ distributions are very similar, as demonstrated in
Fig.~\ref{FIG:YGTEVEX}b. This is not surprising, since the contributions
from the $2\rightarrow 3$ processes are suppressed in the 0-jet
configuration. Apart from the photon bremsstrahlung term which
contributes negligibly for the photon isolation cut we impose [see
Eq.~(\ref{EQ:ISOL})], all $2\rightarrow 2$ contributions preserve the
radiation zero.

The decomposition of the inclusive NLO photon $p_T^{}$ distribution at the
SSC into 0-jet and 1-jet fractions is shown in
Fig.~\ref{FIG:PTGSSCEX}. For transverse momenta close to the
minimum $p_T^{} (\gamma)$ threshold,
the 0-jet and 1-jet rates are approximately equal. At high $p_T^{}(\gamma)
$, the 1-jet cross section completely dominates. In
Fig.~\ref{FIG:PTGSSCEX}b, the NLO 0-jet photon
$p_T^{}$ distribution is compared to the photon transverse momentum
distribution in the Born approximation. Although the QCD corrections for
$W\gamma+0$~jet production are much smaller than in the inclusive
reaction $pp\to W\gamma+X$, they are still sizable. At small $p_T^{}$, the
${\cal O}(\alpha_s)$ corrections to $W\gamma +0$~jet production increase
the cross section by about a factor~2 for the parameters used, whereas
the QCD corrected cross section is somewhat smaller than the result
obtained in the Born approximation at high photon transverse momenta.

Figure~\ref{FIG:YGSSCEX}a shows the inclusive NLO, the ${\cal
O}(\alpha_s)$ $W\gamma+0$~jet, and the LO $W\gamma+1$~jet $|y^*_{\gamma}|$
distributions for $pp$ collisions at $\sqrt{s} = 40$~TeV.
At small rapidities, the 1-jet
channel contributes about 60\% to the inclusive NLO cross section; for
$|y^*_\gamma|\geq 1.4$ the 0-jet and 1-jet cross sections are
approximately equal. In Fig.~\ref{FIG:YGSSCEX}b we compare the NLO 0-jet
result with the prediction obtained in the Born approximation. For the
jet definition of Eq.~(\ref{EQ:SSCJET}), the QCD corrections
to the $W \gamma + 0$~jet cross section completely fill the dip
at $|y^*_\gamma|=0$. For $|y^*_\gamma|\leq 0.7$, the NLO 0-jet cross
section is almost completely flat. Even if the jet defining $p_T^{}$
threshold is reduced to 30~GeV, the radiation zero is still completely
obscured by the QCD corrections. For the reduced $p_T^{}$ threshold, the
NLO 0-jet $|y^*_\gamma|$ distribution almost coincides with the result
obtained in the Born approximation for $|y^*_\gamma|>1$ and is
practically constant for $|y^*_\gamma|<0.7$.

One of the motivations for performing NLO calculations is that the
results often show a less dramatic dependence on the renormalization
and factorization scale than the LO result. Figure~\ref{FIG:QSCALE}
shows the scale dependence of the Born, the inclusive NLO, the ${\cal
O}(\alpha_s)$ 0-jet exclusive, and the 1-jet exclusive cross sections for the
Tevatron, LHC, and SSC center of mass energies. To obtain the cross
section at LHC energies, the same cuts and jet definition as
for the SSC have been imposed. The total cross section for the reaction
$p\,p\hskip-7pt\hbox{$^{^{(\!-\!)}}$} \rightarrow W^+\gamma + X
\rightarrow e^+ \nu \gamma + X$ is plotted versus the scale $Q$.
The factorization scale $M^2$ and the renormalization scale $\mu^2$
have both been set equal to $Q^2$.

The scale dependence of the Born cross section enters only through the
$Q^2$ dependence of the parton distribution functions. The qualitative
differences between the results at the Tevatron and the supercolliders
are due to the differences between $p \bar p$ versus $pp$ scattering
and the ranges of the $x$-values probed.  At the Tevatron, $W\gamma$
production in $p \bar p$ collisions is dominated by valence quark
interactions.  The valence quark distributions decrease slightly with $Q^2$
for the $x$-values probed at the Tevatron.  On the other
hand, at the LHC and SSC, sea quark interactions dominate in the $pp$
process and smaller $x$-values are probed.  The sea quark
distributions increase with $Q^2$ for the $x$-values probed at the
LHC and SSC.  Thus the Born cross section decreases slightly with
$Q^2$ at the Tevatron but increases with $Q^2$ at the LHC and SSC.
The relative stability of the Born cross section at the Tevatron is
accidental and depends on the cuts. For a larger $p_T^{}(\gamma)$ cut,
the Born cross section varies more strongly with $Q$.

The scale dependence of the 1-jet exclusive cross section enters via
the parton distribution functions and the running coupling $\alpha_s(Q^2)$.
Note that the 1-jet exclusive cross section is calculated only to
lowest order and thus exhibits a considerable scale dependence.
The dependence on $Q$ here is dominated by the scale dependence
of $\alpha_s(Q^2)$ which is a decreasing function of $Q^2$.
At the NLO level, the $Q$ dependence enters not only via the
parton distribution functions and the running coupling $\alpha_s(Q^2)$,
but also through explicit factorization scale dependence in the
order $\alpha_s(Q^2)$ correction terms [see Eq.~(\ref{EQ:NEWNLO})].
The NLO 0-jet exclusive cross section is almost independent of the scale
$Q$. Here, the scale dependence of the
parton distribution functions is compensated by that of $\alpha_s(Q^2)$
and the explicit factorization scale dependence in the correction terms.
The $Q$ dependence of the inclusive NLO cross section is dominated by the
1-jet exclusive component and is significantly larger than that of the NLO
0-jet cross section.
(The slight differences between the scale dependencies shown here and in
Ref.~\cite{WGAMMA} are due to the different cuts on the final state particles.)

The results obtained for the NLO exclusive $W\gamma +0$~jet and the
LO exclusive $W\gamma+1$~jet differential cross
sections depend explicitly on the jet definition.  Only the
inclusive NLO distributions are independent of the jet definition.
The sensitivity of the NLO $W^+\gamma +0$~jet
differential cross section to the jet defining $p_T^{}$ threshold is
investigated in Fig.~\ref{FIG:LAST}a where we compare the photon
transverse momentum
distribution obtained in the Born approximation (solid line) with the
$p_T^{}(\gamma)$ spectrum of the NLO $W^+\gamma +0$~jet process for two
different jet
definitions at the SSC. The dashed line shows the result obtained using the
definition of Eq.~(\ref{EQ:SSCJET}). The dotted line displays the result if the
$p_T^{}(j)$ threshold is lowered to 30~GeV. In this case, the $p_T^{}(\gamma)$
differential cross section is approximately 30\% smaller than the result
obtained for a 50~GeV $p_T^{}(j)$ threshold. Present studies~\cite{RSDC}
suggest that jets with $p_T^{}>50$~GeV can be identified at the SSC without
problems, whereas it will be difficult to reconstruct a jet with a transverse
momentum smaller than about 30~GeV. The dashed and dotted lines in
Fig.~\ref{FIG:LAST}a therefore represent the typical uncertainties
in the NLO $W\gamma +0$~jet cross section originating from the jet
definition at the SSC. Qualitatively similar results are obtained for
the Tevatron. The jet transverse momentum threshold can also not be lowered to
arbitrarily small values in our calculation for theoretical reasons.
For transverse momenta below 5~GeV (20~GeV) at the Tevatron
(SSC), soft gluon resummation effects are expected to significantly change
the jet $p_T^{}$ distribution~\cite{RESUM}. These effects are not included
in our calculation.

In Fig.~\ref{FIG:QSCALE} we illustrated the dependence of the total
cross section on the factorization and renormalization scale $Q$. The
total cross section, however, only poorly reflects
the scale dependence of the differential cross section. In
Fig.~\ref{FIG:LAST}b we investigate the $Q$ dependence of the $p_T^{}(\gamma)$
differential cross section at the SSC for $W^+\gamma +0$~jet production at NLO,
using the jet definition of Eq.~(\ref{EQ:SSCJET}).
Results are shown for $Q^2=M^2_{W\gamma}$ (dashed line)
and $Q^2=M_W^2$ (dotted line). The result obtained
for the NLO $W^+ \gamma + 0$~jet $p_T^{}(\gamma)$
distribution is almost independent of the scale over the
whole range of $p_T^{}(\gamma)$ shown. In contrast, the $p_T^{}(\gamma)$
differential cross section obtained in the Born approximation displays a
slight change in shape if $Q^2$ is changed
from $M^2_{W\gamma}$ (solid line) to $M_W^2$ (dot-dashed line).
The variation of the shape of the photon $p_T^{}$
distribution with $Q^2$ in the Born approximation is somewhat more
pronounced at Tevatron energies.

The results shown in Figs.~\ref{FIG:PTGTEVEX} --~\ref{FIG:YGSSCEX}
suggest that the size of the ${\cal
O}(\alpha_s)$ QCD corrections can be significantly reduced by vetoing
hard jets in the central rapidity region, {\it i.e.}, by imposing a
``zero jet'' requirement and considering the $W\gamma+0$~jet channel
only. A zero jet cut for example has been imposed in the CDF measurement
of the ratio of $W$ to $Z$ cross sections~\cite{RATIO} and the $W$ mass
measurement~\cite{WMASS}. Figure~\ref{FIG:PTGBORNZERO} demonstrates that
a jet veto to a large extent restores the sensitivity to
anomalous $WW\gamma$ couplings lost in the inclusive NLO case at the SSC.
Vetoing against jets with $p_T^{}(j)>50$~GeV and
$|\eta(j)|<3$, the ${\cal O}(\alpha_s)$ QCD corrections affect the shape
in the SM case, as well as for anomalous couplings, only
modestly. For non-standard $WW\gamma$ couplings, the shape is changed
in a significant way for photon transverse momenta below 400~GeV only.

\subsection{Sensitivity Limits}

As we have demonstrated so far, ${\cal O}(\alpha_s)$ QCD corrections
significantly affect $W\gamma$ production at hadron colliders and may
reduce the
sensitivity to anomalous $WW\gamma$ couplings substantially unless a
jet veto is imposed. We now want to make this statement more
quantitative by comparing the sensitivity limits for $\Delta\kappa_0$
and $\lambda_0$ achievable at the
Tevatron and SSC for $W\gamma$ production in the Born approximation with
the bounds obtained from the inclusive NLO $W\gamma+X$ and the
exclusive NLO $W\gamma +
0$~jet calculation. To derive $1\sigma$ and $2\sigma$ (68\% and 95\%
confidence level) limits we use the $p_T^{}(\gamma)$ distribution and
assume an integrated luminosity of 100~pb$^{-1}$ at the Tevatron and
$10^4$~pb$^{-1}$ at the SSC. In the Born approximation, the photon
transverse momentum distribution in general yields the best sensitivity
bounds. Furthermore, we use the cuts summarized in
Section~IIIB and the jet definitions in Eqs.~(\ref{EQ:TEVJET})
and~(\ref{EQ:SSCJET}). Only $W\rightarrow e\nu$ decays are taken
into account in our analysis. To extract limits at the Tevatron, we shall
sum over both $W$ charges. For the SSC, we consider only $W^+\gamma$
production. Interference effects between $\Delta\kappa_0$ and
$\lambda_0$ are fully incorporated in our analysis.

The statistical significance is calculated by splitting the $p_T^{}(\gamma)
$ distribution into 8 (5) bins at the SSC (Tevatron). In each bin the
Poisson statistics are approximated by a Gaussian distribution. In order
to achieve a sizable counting rate in each bin, all events with
$p_T^{}(\gamma)> 450$~GeV (30~GeV) at the SSC (Tevatron) are collected in a
single bin. This guarantees that a high statistical significance cannot
arise from a single event at large transverse momentum, where the
SM predicts, say, only 0.01 events. In order to derive
realistic limits we allow for a normalization uncertainty of 50\% in the
SM cross section. Background contributions are
ignored in our derivation of sensitivity bounds.

Our results are summarized in Table~1. The limits for $\Delta\kappa_0$
apply for arbitrary values of $\lambda_0$ and vice versa. At the
Tevatron, QCD corrections
only slightly influence the sensitivities which can be achieved with
100~pb$^{-1}$. The inclusive NLO $W^\pm\gamma+X$ and the exclusive NLO
$W^\pm\gamma+0$~jet limits are virtually identical. From the discussion
in Section~IIID and~IIIE one would expect that the sensitivity
limits from the inclusive NLO $W^\pm\gamma+X$ cross section are somewhat
worse than those obtained using the Born approximation. Table~1 shows
that this is not the case. This result can be easily understood by
noting that only about 11~events with $p_T^{}(\gamma)>30$~GeV are expected
in the SM, including ${\cal O}(\alpha_S)$ QCD corrections, for
100~pb$^{-1}$. Due to the small number of events, the reduced sensitivity to
anomalous $WW\gamma$ couplings originating from the shape change induced
by the ${\cal O}(\alpha_s)$ corrections at large photon transverse momenta is
not reflected in the bounds which can be achieved for the
anomalous couplings. The slight improvement with respect to the limits
obtained using the Born approximation is due to the increased cross
section in the inclusive NLO case.

At SSC energies, the situation changes quite drastically. Inclusive
NLO QCD corrections reduce the sensitivity to $\Delta\kappa_0$
($\lambda_0$) by a factor $\sim 1.7$ ($\sim 2.1$), although the
inclusive ${\cal O}(\alpha_s)$ corrections increase the total cross
section by more than a factor~3. Furthermore,
interference effects between the SM and anomalous terms in
the helicity amplitudes considerably increase when inclusive NLO QCD
corrections are taken into account. As a result, the bounds in the
inclusive NLO case depend significantly on the sign of the anomalous
coupling, in contrast to the limits obtained in the Born approximation.
The increase of these interference effects is due to the logarithmic
enhancement factor which is present
in the SM quark-gluon fusion term at large photon transverse momenta.

A large portion of the
sensitivity lost in the inclusive NLO case can be regained if a jet veto
is imposed. The NLO
$W^+\gamma +0$~jet limits are typically 10 -- 20\% weaker than
those obtained in the Born approximation and depend only marginally
on the jet definition criteria. In Section~IIIE we found that the NLO
$W\gamma+0$~jet differential cross section is more stable to variations
of the factorization scale $Q^2$ than the Born and inclusive NLO
$W\gamma+X$ cross sections (see Figs.~\ref{FIG:QSCALE} and~\ref{FIG:LAST}a).
The systematic errors which originate from the choice of $Q^2$ thus
will be smaller for bounds derived from the NLO $W\gamma+0$~jet
differential cross section than those for limits obtained from the inclusive
NLO $W\gamma+X$ or the Born cross section. Note that the results shown in
Table~1 automatically imply that $W\gamma+1$~jet production, with a high
transverse momentum jet, will be less sensitive to anomalous $WW\gamma$
couplings than $W\gamma+0$~jet production.

The bounds shown in Table~1 have been derived for a dipole form factor
($n=2$) with a scale of $\Lambda=1$~TeV. At Tevatron energies, the
sensitivities achievable are insensitive to the exact form and the
scale of the form factor (for $\Lambda>400$~GeV). At the SSC, the
situation is different and
the sensitivity bounds depend significantly on the value chosen for
$\Lambda$~\cite{BAURZEP}. For $\Lambda=5$~TeV, for example, the limits
of Table~1b improve by a factor $\sim 2.5$ ($\sim 2$) for
$\Delta\kappa_0$ ($\lambda_0$). The bounds in the inclusive NLO case are
again weaker by up to a factor of~2 compared to those obtained in the
Born approximation. If a jet veto is imposed, the sensitivities
achievable are very similar to those found in the Born approximation. The
usefulness of the zero jet requirement, thus, does not depend on details of the
form factor assumed for the non-standard $WW\gamma$ couplings.

In Table~1, we have shown sensitivity limits only for the Tevatron and
SSC. At LHC energies, the situation is very similar to that encountered
at the SSC. For an integrated luminosity of $10^4$~pb$^{-1}$, the
sensitivities which can be achieved at the LHC are about a factor 1.5
worse than those expected for the SSC.

The bounds displayed in Table~1 are quite conservative. If
$W\rightarrow\mu\nu$ decays and, at the SSC, $W^-\gamma + X$ production
are included, the limits can easily be improved by 20 -- 40\%.
Further improvements may result from using more powerful statistical
tools than the simple $\chi^2$ test we performed. Our results,
however, clearly demonstrate the advantage of a jet veto to probe the
structure of the $WW\gamma$ vertex in $W\gamma$ production at hadron
supercolliders.

\newpage
\section{SUMMARY}

$W\gamma$ production in hadronic collisions provides an opportunity to
probe the structure of the $WW\gamma$ vertex in a direct and
essentially model
independent way. Previous studies of $p\,p\hskip-7pt\hbox{$^{^{(\!-\!)
}}$} \rightarrow W^{\pm}\gamma$~\cite{BAURZEP,BAURBERGER} have been
based on leading order calculations. In this paper we have presented
an ${\cal O}(\alpha_s)$ calculation of the reaction
$p\,p\hskip-7pt\hbox{$^{^{(\!-\!)}}$} \rightarrow W^{\pm}\gamma + X
\rightarrow \ell^\pm \nu \gamma + X$ for general, $CP$ conserving,
$WW\gamma$ couplings, using a combination of analytic and Monte Carlo
integration techniques. The $W\rightarrow \ell\nu$ decay has been included
in the narrow width approximation in our calculation. In this
approximation, diagrams in which the photon is
radiated off the final state lepton line are not necessary to maintain
electromagnetic gauge invariance. For suitable cuts these diagrams can
thus be ignored, which considerably simplifies the calculation.
$W$ decay spin correlations are correctly taken into account in our
approach, except in the finite virtual contribution. The finite virtual
correction term contributes only at the per cent level to the total cross
section and $W$ decay spin correlations can thus be safely ignored here.

The photon $p_T^{}$ differential cross section
is very sensitive to non-standard $WW\gamma$ couplings. We found that
QCD corrections significantly change the shape of this distribution.
This shape change is due to a combination of destructive interference in the
$W\gamma$ Born subprocess and a logarithmic enhancement factor in
the $qg$ and $\bar q g$ real emission subprocesses.
The destructive interference suppresses the size of the $W\gamma$ Born
cross section and is also responsible for the radiation amplitude zero.
The logarithmic enhancement factor originates in the high $p_T^{} (\gamma)$
region of phase space where the photon is balanced by a high $p_T^{}$
quark which radiates a soft $W$ boson.
The logarithmic enhancement factor and the large gluon density
make the ${\cal O} (\alpha_s)$ corrections large at high $p_T^{} (\gamma)$,
especially when the center of mass energy is large.

Since the Feynman diagrams responsible for the enhancement at large
photon transverse momenta do not involve the $WW\gamma$ vertex,
inclusive NLO QCD corrections to $W^\pm\gamma$ production tend to reduce the
sensitivity to non-standard couplings. At the Tevatron, for an
integrated luminosity of 100~pb$^{-1}$, this effect is overwhelmed by
the increase in cross section induced by the QCD corrections. Due to the
very large quark-gluon luminosity at the LHC and SSC, however, one
expects that the sensitivity bounds which can be achieved at those
machines are reduced by up to a factor~2 (see Table~1).

The size of the QCD corrections at large photon transverse momenta may
be reduced substantially and a large fraction of the sensitivity to
anomalous $WW\gamma$ couplings which was lost at SSC and LHC energies
may be regained by imposing a jet veto, {\it i.e.}, by considering the
exclusive $W\gamma+0$~jet channel instead of inclusive $W\gamma+X$
production. Such a ``zero-jet'' requirement may be also very helpful
to suppress the background from $t\bar t\gamma$ production~\cite{MAINA}
at the LHC and SSC. Furthermore, we found that the
dependence of the NLO $W\gamma+0$~jet cross section on the factorization
scale $Q^2$ is significantly reduced compared to that of the inclusive
NLO $W\gamma+X$ cross section. Uncertainties which
originate from the variation of $Q^2$ thus will be smaller for
sensitivity bounds obtained from the $W\gamma+0$~jet channel than those
for limits derived from the inclusive NLO $W\gamma+X$ cross section.

Although the magnitude of the QCD corrections at
SSC energies is significantly reduced if a jet veto is imposed, the
residual NLO corrections to $W\gamma+0$~jet production are still quite
large, in particular for small values of $p_T^{}(\gamma)$, and cannot be
ignored. This also means that in order to complete our understanding of QCD
corrections in $W\gamma$ production, a full calculation of the ${\cal
O}(\alpha_s^2)$ corrections at SSC energies will be necessary. These
corrections have recently
been calculated in Ref.~\cite{MENDOZA} in the soft-plus-virtual gluon
approximation for CERN $p\bar p$ collider and Tevatron energies. Jet
vetoing may also be
useful to reduce the size of the ${\cal O}(\alpha_s^2)$ corrections.

Our results show that NLO QCD corrections only slightly
influence the sensitivity limits which can be achieved at the Tevatron.
Nevertheless, it will be important to take these corrections into
account when extracting information on the structure of the $WW\gamma$
vertex, in order to reduce systematic and theoretical errors. At the LHC
and SSC it will be absolutely necessary to take into account the effects
of higher order QCD corrections when experimental data and theoretical
predictions for $W\gamma$ production are compared.

%
\acknowledgements

We would like to thank S.~Errede, E.~Laenen, J.~Smith, G.~Valencia,
and D.~Zeppenfeld for stimulating discussions. Two of us (UB and JO)
wish to thank the Fermilab Theory Group and the Institute for
Elementary Particle Physics Research at the University of Wisconsin--Madison
for their warm hospitality during various stages of this work. This
work was supported
in part by the UK Science and Engineering Research Council and the U.~S.
Department of Energy under Contract No.~DE-FG05-87ER40319. T.~Han was supported
in part by the Texas National Research Laboratory Commission under Award
No.~FCFY9116.

\newpage
%
%
\appendix{\ \ PHOTON BREMSSTRAHLUNG}

The photon bremsstrahlung contribution to $W\gamma$ production and decay
is calculated by convoluting the ${\cal O}(\alpha_s)$ hard scattering
subprocess cross section for $W$ production and decay with the
appropriate parton distribution and fragmentation functions:
\begin{eqnarray}
\sigma_{\hbox{\scriptsize brem}}^{\phantom{NLO}} &=& \sum_{a,b,c} \> \int
G_{a/A} (x_a,M^2) \, G_{b/B} (x_b,M^2) \, D_{\gamma/c} (z_c,M^2)
\label{EQ:SINGLEBREM} \\
\noalign{\vskip 5pt}
&&\qquad\qquad \times {d \hat \sigma \over dv}
(a b \to W c \to \ell \nu c) \, dx_a \, dx_b \, dz_c \, dv \>.
\nonumber
\end{eqnarray}
The squared matrix element for the subprocess
$q_1(p_1) + \bar q_2(p_2) \to W g \to \ell(p_3) + \nu(p_4) +  g(p_5)$ is
\FL
\begin{eqnarray}
\Bigl|{\cal M}(q_1 \bar q_2 \to W g \to \ell \nu g) \Bigr|^2 &=&
{2^9 \, \pi^3 \, \alpha_s \, \alpha^2 \over x^2_{\hbox{\scriptsize w}} } \,
{s_{34} \over t_{15} t_{25}} \,
{ \Bigl[ t_{14}^2 + t_{23}^2 \Bigr] \over
  \Bigl[ (s_{34} - M_W^2)^2 + (\Gamma_W M_W)^2 \Bigr] } \>,
\end{eqnarray}
where $s_{ij} = (p_i + p_j)^2$, $t_{ij}= (p_i - p_j)^2$,
and $\Gamma_W$ is the total width of the $W$-boson;
spin and color averages are not included. The squared matrix
elements for the subprocesses $q_1 g \to W q_2 \to \ell \nu q_2$
and $g \bar q_2 \to W \bar q_1 \to \ell \nu \bar q_1$ are obtained by crossing
$p_2 \leftrightarrow - p_5$ and $p_1 \leftrightarrow -p_5$, respectively,
and introducing an overall minus sign.  If a photon isolation cut of the
type discussed in Section~IIIB is included, then the range of $z$ is
reduced from $0 \le z \le 1$ to
$1/(1+\epsilon_{\hbox{\scriptsize h}}) \le z \le 1$.

The LO bremsstrahlung cross section is obtained by using leading-log
fragmentation functions.  The numerical work in this paper was done
using the parameterizations of Ref.~\cite{DUKE} for the LO
fragmentation functions:
\FL
\begin{eqnarray}
z D^{\hbox{\scriptsize LO}}_{\gamma /q} (z,Q^2) &=&
 F \Biggl[ {Q_q^2 (2.21-1.28z+1.29z^2)z^{0.049} \over
 1 - 1.63 \ln(1-z) } + 0.0020 (1-z)^{2.0} z^{-1.54} \Biggr] ,\quad
\label{EQ:PHOTONFRAGQ} \\
\noalign{\vskip 5pt}
z D^{\hbox{\scriptsize LO}}_{\gamma /g} (z,Q^2) &=&
{0.194\over 8} \, F \, (1-z)^{1.03} \, z^{-0.97} ,
\label{EQ:PHOTONFRAGG}
\end{eqnarray}
where $Q_q$ is the electric charge of the quark $q$
(in units of the proton charge $e$),
$F = (\alpha / 2 \pi) \ln (Q^2 / \Lambda^2)$,
and $\Lambda = \Lambda_4$. Since $\alpha_s(Q^2) = 12\pi/[(33-2N_F)
\ln(Q^2/\Lambda^2)]$, these fragmentation functions
are proportional to $\alpha/\alpha_s$.

The logarithmic growth of the fragmentation functions arises from an
integration over the transverse momentum of the photon with respect to
the quark.  The upper limit for this integration has been taken to be
the typical hard scattering momentum scale $Q^2$.  The divergence
associated with the lower limit has been regulated by using the QCD
scale parameter $\Lambda$ as an  infrared cutoff.  Details on
the derivation of these fragmentation functions can be found in
Refs.~\cite{DUKE} and~\cite{JFO}.

At the next-to-leading-order there are collinear singularities
associated with final state bremsstrahlung which must be factorized
and absorbed into  fragmentation functions.  This will modify the
leading-order quark fragmentation functions such that
\begin{eqnarray}
D^{\hbox{\scriptsize NLO}}_{\gamma/q} (z) = D^{{\hbox{\scriptsize
LO}}}_{\gamma/q} (z)
+ {\alpha \over 2\pi} \, Q_q^2 \,
\biggl[ \Bigl\{ {1 + (1-z)^2 \over z} \Bigr\}
\, \ln \Bigl\{ z(1-z) \delta_c {s_{12} \over M^2} \Bigr\} + z \biggr] \>.
\label{EQ:NLOFRAG}
\end{eqnarray}
Here $\delta_c$ is the collinear cutoff parameter and $M^2$ is the
factorization scale. The new term is the
remnant of the collinear singularity after the factorization process
has been performed. The gluon fragmentation function is unchanged.
\newpage
%

\appendix{\ \ HARD COLLINEAR CORRECTIONS}

The real emission processes, {\it e.g.},
$q_1(p_1) + \bar q_2(p_2) \to W \gamma g \to \ell(p_3)
+ \nu(p_4) + \gamma(p_5) + g(p_6)$, have hard collinear
singularities when $t_{16} \rightarrow 0$ or $t_{26} \rightarrow 0$.
These singularities must be factorized and absorbed into the initial
state parton distribution functions. After the factorization is
performed, the contribution from the remnants
of the hard collinear singularities has the form
\widetext
\begin{eqnarray}
\sigma^{\hbox{\scriptsize hc}} &=& \sum_{q_1,\bar q_2} \int
 {\alpha_s \over 2 \pi} \,
 {d\hat\sigma^{\hbox{\scriptsize Born}} \over dv}
 (q_1 \bar q_2 \rightarrow W\gamma \to \ell \nu \gamma) \,
dv\,dx_1\,dx_2 \label{EQ:HARDCOL} \\
\noalign{\vskip 5pt}
& &\quad \times \Biggl[ \phantom{+}
G_{q_1/p}(x_1,M^2) \int\limits_{x_2}^{1 - \delta_s} \, {dz \over z} \,
G_{\bar q_2/p\hskip-7pt\hbox{$^{^{(\!-\!)}}$}} \Bigl({x_2\over z},
M^2\Bigr) \, \tilde P_{qq}(z)  \nonumber \\
\noalign{\vskip 5pt}
& &\quad \phantom{\times \Biggl[}
+ G_{q_1/p}(x_1,M^2) \int\limits_{x_2}^{1} \, {dz \over z} \,
  G_{g/p} \Bigl({x_2\over z},M^2\Bigr) \, \tilde P_{qg}(z) \nonumber \\
\noalign{\vskip 5pt}
& &\quad \phantom{\times \Biggl[ }
 + G_{\bar q_2/p\hskip-7pt\hbox{$^{^{(\!-\!)}}$}}(x_2,M^2) \int\limits_
{x_1}^{1 - \delta_s} \, {dz \over z} \,
   G_{q_1/p} \Bigl({x_1\over z},M^2\Bigr) \, \tilde P_{qq}(z)
 \nonumber \\
\noalign{\vskip 5pt}
& &\quad \phantom{\times \Biggl[ }
 + G_{\bar q_2/p\hskip-7pt\hbox{$^{^{(\!-\!)}}$}}(x_2,M^2) \int\limits_
{x_1}^{1} \,  {dz \over z} \,
  G_{g/p} \Bigl( {x_1\over z},M^2\Bigr) \, \tilde P_{qg}(z) \Biggr] \>,
\nonumber
\end{eqnarray}
\narrowtext
with
\FL
\begin{eqnarray}
\tilde P_{ij}(z) \equiv P_{ij}(z) \, \ln \biggl( {1-z \over z} \,
 \delta_c \, {s_{12} \over M^2} \biggr)
 - P_{ij}^{\prime}(z) - \lambda_{FC} \, F_{ij}(z) \>.
\label{EQ:AP}
\end{eqnarray}
The Altarelli-Parisi splitting functions in $N = 4 - 2 \epsilon$
dimensions for $0 < z < 1$ are
\begin{eqnarray}
P_{qq} (z,\epsilon) &=& C_F \, \left[ {1 + z^2 \over 1 - z} -
 \epsilon (1 - z) \right] \>, \\
\noalign{\vskip 10pt}
P_{qg} (z,\epsilon) &=& {1 \over 2(1 - \epsilon)} \,
 \Biggl[z^2 + (1-z)^2 - \epsilon \Biggr] \>,
\end{eqnarray}
and can be written
\begin{eqnarray}
P_{ij} (z,\epsilon) = P_{ij} (z) + \epsilon P_{ij}^{\prime} (z) \>,
\end{eqnarray}
which defines the $P_{ij}^{\prime}$ functions. The functions $F_{qq}$
and $F_{qg}$ depend on the choice of factorization  convention and the
parameter $\lambda_{FC}$ specifies the factorization convention;
$\lambda_{FC} = 0$ for the universal
(Modified Minimal Subtraction ${\rm \overline{MS}}$~\cite{MSBAR})
convention and $\lambda_{FC} = 1$ for the physical (Deep Inelastic
Scattering DIS) convention. For the physical convention the
factorization functions are
\begin{eqnarray}
F_{qq} (z) &=& C_F \, \left[ {1+z^2 \over 1-z} \>
\ln \left({1-z \over z} \right)
- {3 \over 2} \, {1 \over 1-z} + 2 z + 3 \right] , \\
\noalign{\vskip 10pt}
F_{qg} (z) &=& {1\over 2} \left[ \left\{ z^2 + (1-z)^2 \right\}
\ln \left( {1-z \over z} \right) + 8z(1-z) -1 \right] .
\label{EQ:ORIGFQG}
\end{eqnarray}
The transformation between the ${\rm \overline{MS}}$
and DIS schemes is discussed in Ref.~\cite{OWENSTUNG}.
The parameter $M^2$ is the factorization scale which must be specified
in the process of  factorizing the collinear singularity. Basically,
it determines how much of the collinear  term is absorbed into the
various parton distributions.

\newpage
%
%

%
\newpage
%
\widetext
\begin{table}
\caption{Sensitivities achievable at the $1\sigma$ and $2\sigma$ confidence
levels (CL) for the anomalous $WW\gamma$ couplings $\Delta\kappa_0$ and
$\lambda_0$ in $p\bar p\rightarrow W^\pm\gamma + X\rightarrow
e^\pm\nu\gamma+X$ at the Tevatron and $pp\rightarrow W^+\gamma +
X\rightarrow e^+\nu\gamma+X$ at the SSC. The limits for
$\Delta\kappa_0$ apply for arbitrary values of $\lambda_0$ and vice
versa. For the form factors we use Eqs.~(\ref{EQ:KAPPAFORM})
and~(\ref{EQ:LAMBDAFORM}) with $n=2$ and
$\Lambda=1$~TeV. We assume an integrated luminosity of 100~pb$^{-1}$ at
the Tevatron and $10^4$~pb$^{-1}$ at the SSC. The cuts summarized in
Section~IIIB are imposed. In the NLO 0-jet case we have used the jet
definitions in Eqs.~(\ref{EQ:TEVJET}) and~(\ref{EQ:SSCJET}).}
\label{TABLE1}
\begin{tabular}{ccccc}
\multicolumn{5}{c}{a) Tevatron}\\
coupling
&\multicolumn{1}{c}{CL}
&\multicolumn{1}{c}{Born appr.}
&\multicolumn{1}{c}{incl. NLO}
&\multicolumn{1}{c}{NLO 0-jet}\\
\tableline
$\Delta\kappa_0$ & $2\sigma$ & $\matrix{+1.8 \crc -1.6}$ & $\matrix{+1.7
\crc -1.5}$ & $\matrix{+1.7 \crc -1.5}$ \\ [5.mm]
 & $1\sigma$ & $\matrix{+1.1 \crc -0.8}$ & $\matrix{+1.0 \crc -0.8}$ &
$\matrix{+1.0 \crc -0.8}$ \\ [2.mm]
\tableline
$\lambda_0$ & $2\sigma$ & $\matrix{+0.53 \crc -0.58}$ & $\matrix{+0.50
\crc -0.56}$ & $\matrix{+0.51 \crc -0.57}$ \\ [5.mm]
 & $1\sigma$ & $\matrix{+0.29 \crc -0.35}$ & $\matrix{+0.27 \crc -0.34}$
 & $\matrix{+0.28 \crc -0.34}$ \\ [2.mm]
\tableline
\tableline
\multicolumn{5}{c}{b) SSC}\\
coupling
&\multicolumn{1}{c}{CL}
&\multicolumn{1}{c}{Born appr.}
&\multicolumn{1}{c}{incl. NLO}
&\multicolumn{1}{c}{NLO 0-jet}\\
\tableline
$\Delta\kappa_0$ & $2\sigma$ & $\matrix{+0.33 \crc -0.34}$ &
$\matrix{+0.46 \crc -0.59}$ & $\matrix{+0.37 \crc -0.39}$ \\ [5.mm]
 & $1\sigma$ & $\matrix{+0.18 \crc -0.24}$ & $\matrix{+0.26 \crc -0.39}$ &
$\matrix{+0.22 \crc -0.24}$ \\ [2.mm]
\tableline
$\lambda_0$ & $2\sigma$ & $\matrix{+0.033 \crc -0.029}$ & $\matrix{+0.044
\crc -0.053}$ & $\matrix{+0.033 \crc -0.035}$ \\ [5.mm]
 & $1\sigma$ & $\matrix{+0.022 \crc -0.018}$ & $\matrix{+0.028 \crc -0.038}$
 & $\matrix{+0.020 \crc -0.022}$ \\ [2.mm]
\end{tabular}
\end{table}
\newpage
%
%
\figure{Feynman diagrams for the Born subprocess $q_1\bar
q_2\rightarrow W\gamma \rightarrow \ell \nu \gamma$.
\label{FIG:BORNGRAPHS} }
%
\figure{Feynman diagrams for the virtual subprocess $q_1\bar q_2
\rightarrow W\gamma \rightarrow \ell \nu \gamma$.
Not shown are the diagrams obtained by interchanging the $W$ and $\gamma$.
\label{FIG:VIRTUALGRAPHS} }
%
\figure{Feynman diagrams for the real emission subprocess $q_1\bar
q_2\rightarrow W\gamma g \rightarrow \ell \nu \gamma g$.
Not shown are the diagrams obtained by interchanging the $W$ and $\gamma$.
\label{FIG:REALGRAPHS} }
%
\figure{Feynman rule for the general $WW\gamma$ vertex.
The factor $e$ is the electromagnetic coupling constant and $(Q_1-Q_2)$
is the electric charge of the $W$-boson.  The vertex function
$\Gamma_{\beta \mu \nu}(k,k_1,k_2)$ is given in Eq.~(\ref{EQ:NSMCOUPLINGS}).
\label{FIG:VERTEX} }
%
\figure{Total cross section for $pp \rightarrow W^+ \gamma + X \rightarrow
e^+ \nu_e \gamma + X$ at $\sqrt{s} = 40$~TeV; a) versus $\delta_c$ and
b) versus $\delta_s$. The 3- and 4-body contributions are also shown.
The cuts imposed are summarized in Section~IIIB.
\label{FIG:DELTASC} }
%
\figure{The inclusive differential cross section for the photon transverse
momentum in the reaction
$p \bar p \to W^+ \gamma + X \to e^+ \nu_e \gamma + X$
at $\sqrt{s} = 1.8$~TeV; a) in the Born approximation and b) including
NLO QCD corrections.
The curves are for the SM (solid lines), $\lambda_0 = 0.5$ (dashed lines), and
$\Delta\kappa_0 = 1.0$ (dotted lines).
The cuts imposed are summarized in Section~IIIB.
\label{FIG:PTGTEV} }
%
\figure{The inclusive differential cross section for the reconstructed
$W\gamma$ mass in the reaction
$p \bar p \to W^+ \gamma + X \to e^+ \nu_e \gamma + X$
at $\sqrt{s} = 1.8$~TeV; a) in the Born approximation and b) including
NLO QCD corrections.
The curves are for the SM (solid lines), $\lambda_0 = 0.5$ (dashed lines), and
$\Delta\kappa_0 = 1.0$ (dotted lines).
The cuts imposed are summarized in Section~IIIB.
\label{FIG:MWGTEV} }
%
\figure{The inclusive differential cross section for the photon
rapidity in the reconstructed center of mass frame for the reaction
$p \bar p \to W^+ \gamma + X \to e^+ \nu_e \gamma + X$
at $\sqrt{s} = 1.8$~TeV; a) in the Born approximation and b) including
NLO QCD corrections.
The curves are for the SM (solid lines), $\lambda_0 = 0.5$ (dashed lines), and
$\Delta\kappa_0 = 1.0$ (dotted lines).
The cuts imposed are summarized in Section~IIIB.
\label{FIG:YGTEV} }
%
\figure{The inclusive differential cross section for the photon transverse
momentum in the reaction $p p \to W^+ \gamma + X \to e^+ \nu_e \gamma + X$
at $\sqrt{s} = 40$~TeV; a) in the Born approximation and b) including
NLO QCD corrections.
The curves are for the SM (solid lines), $\lambda_0 = 0.25$ (dashed lines),
and $\Delta\kappa_0 = 1.0$ (dotted lines).
The cuts imposed are summarized in Section~IIIB.
\label{FIG:PTGSSC} }
%
\figure{The inclusive differential cross section for the reconstructed
$W\gamma$ mass in the reaction $pp \to W^+ \gamma + X \to e^+ \nu_e \gamma + X$
at $\sqrt{s} = 40$~TeV; a) in the Born approximation and b) including
NLO QCD corrections.
The curves are for the SM (solid lines), $\lambda_0 = 0.25$ (dashed lines), and
$\Delta\kappa_0 = 1.0$ (dotted lines).
The cuts imposed are summarized in Section~IIIB.
\label{FIG:MWGSSC} }
%
\figure{The inclusive differential cross section for the photon
rapidity in the reconstructed center of mass frame for the reaction
$pp \to W^+ \gamma + X \to e^+ \nu_e \gamma + X$
at $\sqrt{s} = 40$~TeV; a) in the Born approximation and b) including
NLO QCD corrections. The curves are for the SM (solid lines),
$\lambda_0 = 0.25$ (dashed lines), and $\Delta\kappa_0 = 1.0$ (dotted lines).
The cuts imposed are summarized in Section~IIIB.
\label{FIG:YGSSC} }
%
\figure{The differential cross section for the photon transverse
momentum in the reaction
$p \bar p \to W^+ \gamma \to e^+ \nu_e \gamma$ at $\sqrt{s} = 1.8$~TeV
in the SM.
a) The inclusive NLO differential cross section (solid line) is
shown, together with the ${\cal O}(\alpha_s)$ 0-jet (dotted line)
and the (LO) 1-jet (dashed line) exclusive differential cross sections,
using the jet definition in Eq.~(\ref{EQ:TEVJET}).
b) The NLO $W\gamma +0$~jet exclusive differential cross section
(dotted line) is compared with the Born differential cross section (dot
dashed line). The cuts imposed are summarized in Section~IIIB.
\label{FIG:PTGTEVEX} }
%
\figure{The differential cross section for the photon rapidity in the
reconstructed center of mass frame for the reaction
$p \bar p \to W^+ \gamma \to e^+ \nu_e \gamma$ at $\sqrt{s} = 1.8$~TeV
in the SM.
a) The inclusive NLO differential cross section (solid line) is
shown, together with the ${\cal O}(\alpha_s)$ 0-jet (dotted line)
and the (LO) 1-jet (dashed line) exclusive differential cross sections,
using the jet definition in Eq.~(\ref{EQ:TEVJET}).
b) The NLO $W\gamma +0$~jet exclusive differential cross section
(dotted line) is compared with the Born differential cross section (dot
dashed line). The cuts imposed are summarized in Section~IIIB.
\label{FIG:YGTEVEX} }
%
\figure{The differential cross section for the photon transverse
momentum in the reaction
$pp \to W^+ \gamma \to e^+ \nu_e \gamma$ at $\sqrt{s} = 40$~TeV in the
SM.
a) The inclusive NLO differential cross section (solid line) is
shown, together with the ${\cal O}(\alpha_s)$ 0-jet (dotted line)
and the (LO) 1-jet (dashed line) exclusive differential cross sections,
using the jet definition in Eq.~(\ref{EQ:SSCJET}).
b) The NLO $W\gamma +0$~jet exclusive differential cross section
(dotted line) is compared with the Born differential cross section (dot
dashed line). The cuts imposed are summarized in Section~IIIB.
\label{FIG:PTGSSCEX} }
%
\figure{The differential cross section for the photon rapidity in the
reconstructed center of mass frame for the reaction
$pp \to W^+ \gamma \to e^+ \nu_e \gamma$ at $\sqrt{s} = 40$~TeV in the
SM.
a) The inclusive NLO differential cross section (solid line) is
shown, together with the ${\cal O}(\alpha_s)$ 0-jet (dotted line)
and the (LO) 1-jet (dashed line) exclusive differential cross sections,
using the jet definition in Eq.~(\ref{EQ:SSCJET}).
b) The NLO $W\gamma +0$~jet exclusive differential cross section
(dotted line) is compared with the Born differential cross section (dot
dashed line). The cuts imposed are summarized in Section~IIIB.
\label{FIG:YGSSCEX} }
%
\figure{The total cross section for
$p\,p\hskip-7pt\hbox{$^{^{(\!-\!)}}$} \rightarrow W^+ \gamma + X
\rightarrow e^+ \nu_e \gamma + X$ in the SM
versus the scale $Q$; a) at the Tevatron, b) at the LHC, and c) at the
SSC. The curves represent the inclusive NLO (solid lines),
the Born (dot dashed lines), the LO 1-jet exclusive (dashed lines),
and the NLO 0-jet exclusive (dotted lines) cross sections.
The cuts imposed are summarized in Section~IIIB. For the jet definitions,
we have used Eqs.~(\ref{EQ:TEVJET}) and~(\ref{EQ:SSCJET}).
\label{FIG:QSCALE} }
%
\figure{The differential cross section for the photon transverse
momentum in the reaction
$p p \to W^+ \gamma +0~{\rm jet}\to e^+ \nu_e \gamma+0$~jet at
$\sqrt{s} = 40$~TeV in the SM.
a) Comparison of the result obtained in the Born approximation (solid
line) with the NLO prediction for two different $p_T^{}(j)$ thresholds.
b) $Q^2$ dependence of the $p_T^{}(\gamma)$ distribution in the Born
approximation (solid and dot-dashed line) and at next-to-leading order
in $\alpha_s$ (dashed and dotted line).
The cuts imposed are summarized in Section~IIIB.
\label{FIG:LAST} }
%
\figure{The differential cross section for the photon transverse
momentum for the exclusive reaction
$pp \to W^+ \gamma +0~{\rm jet}\to e^+ \nu_e \gamma+0$~jet at $\sqrt{s}
= 40$~TeV; a) in the Born approximation and b) including NLO QCD corrections.
The curves are for the SM (solid lines), $\lambda_0 = 0.25$ (dashed lines),
and $\Delta\kappa_0 = 1.0$ (dotted lines).
The cuts imposed are summarized in Section~IIIB. For the jet definition,
we have used Eq.~(\ref{EQ:SSCJET}).
\label{FIG:PTGBORNZERO} }
%
%
\end{narrowtext}
\end{document}